# Vortex Loops In The 3-d XY Model

Arjan Hulsebos[†]

DAMTP
Chadwick Tower
University of Liverpool
P.O. Box 147
Liverpool, L69 3BX
United Kingdom

and

FB 8[‡]
Bergische Universität Gesamthochschule Wuppertal
Gaußstraße 20
42119 Wuppertal
BRD

## Abstract

We study the properties of vortex loops in the 3-d XY model. We find that the phase transition in this model is driven by the condensation of short loops into percolating loops.

---

[†]Internet: `arjanh@wpts0.physik.uni-wuppertal.de`
[‡] Present address.

# 1 Introduction

The analogy between the 2-d XY model and its 3-d counterpart in the sense that both models possess a phase transition at finite critical coupling $\beta_c$, driven by vortices and anti-vortices, may one lead to suspect that the behavior of the vortices and anti-vortices in the 3-d XY model is analogous to the 2-d XY model.

The generally accepted picture of the phase transition in the 2-d XY model is that for couplings $\beta \lesssim \beta_c$, the vortices and anti-vortices are present with finite density, and behave effectively as free particles. As the coupling approaches the critical coupling, the vortices and anti-vortices start to form bound pairs, and may disappear from the system by annihilation, thereby decreasing the vortex density, or vorticity. For couplings $\beta \gtrsim \beta_c$, the system is described by spin waves, while vortices and anti-vortices are only present as tightly bound pairs at (very) low density [1].

If we impose periodic boundary conditions, we will find that in the 3-d case the vortices and anti-vortices will have to form closed loops. The importance of vortex loops on the phase transition of the 3-d XY model has been noted by [2]. The intuitive picture of the phase transition in the 3-d XY model is that for couplings $\beta > \beta_c$, there will be few small loops present in the system. As the coupling $\beta$ is lowered, several possibilities arise. Either the number of loops will increase while the length of the loops remains (roughly) constant, or the length of the loops will increase, while this time the number of loops reamins constant. Also an increase in both is possible.

We will study this picture of the phase transition by measuring properties of these vortex loops. A preliminary account of this work is given in [3].

This paper is divided into sections as follows. In section 2, we will describe the 3-d XY model, and introduce the vortex loops. After discussing some of their properties, we will define the relevant vortex loop correlation functions and distributions. Section 3 will be devoted to the presentation and discussion of the results. In section 4, we will recapitulate these results and give a refined description of the phase transition.

# 2 The 3-d XY Model

The action for the 3-d XY model is given by

$$S = \beta \sum_{x,\mu} \cos(\theta_{x+\hat{\mu}} - \theta_x). \tag{1}$$

The quantity

$$k_{\mu\nu}(x) = \frac{1}{2\pi}\Big\{[\theta_{x+\hat{\mu}} - \theta_x] + [\theta_{x+\hat{\mu}+\hat{\nu}} - \theta_{x+\hat{\mu}}] + [\theta_{x+\hat{\nu}} - \theta_{x+\hat{\mu}+\hat{\nu}}] + [\theta_x - \theta_{x+\hat{\nu}}]\Big\}, \quad \mu < \nu, \tag{2}$$

where $[\ldots]$ denotes the restriction to the interval $<-\pi, \pi]$, yields the vortex number on the corresponding plaquette $P_{\mu\nu}(x)$. To study the properties of these vortices, we perform a duality transformation on $k$ by defining [4]

$$j_\mu(\tilde{x} - \hat{\mu}) = \frac{1}{2}\epsilon_{\mu\nu\rho} k_{\nu\rho}(x), \tag{3}$$

where $\tilde{x} = x + \frac{1}{2}\hat{\mu} + \frac{1}{2}\hat{\nu} + \frac{1}{2}\hat{\rho}$.

The links on the dual lattice have to satisfy the following relation:

$$\partial'_\mu j_\mu(\tilde{x}) = 0 \quad \forall \tilde{x}, \tag{4}$$
$$\partial'_\mu f(\tilde{x}) \equiv f(\tilde{x}) - f(\tilde{x} - \hat{\mu}).$$

This means that every point $\tilde{x}$ on the dual lattice is visited by an even number of non-zero links $j$. On a finite lattice, this implies that we find closed loops, or clusters of closed loops.

Consider now the following spin configuration on an elementary cube:

$$\theta_x = \theta_{x+\hat{2}} = 0,$$
$$\theta_{x+\hat{1}} = \theta_{x+\hat{3}} = -\tfrac{1}{4}\pi,$$
$$\theta_{x+\hat{2}+\hat{3}} = \theta_{x+\hat{1}+\hat{2}} = \tfrac{1}{4}\pi,$$
$$\theta_{x+\hat{1}+\hat{3}} = -\tfrac{3}{4}\pi,$$
$$\theta_{x+\hat{1}+\hat{2}+\hat{3}} = \tfrac{3}{4}\pi. \tag{5}$$

See also figure 1. This configuration has a vortex $k = 1$ on the right face, and an anti-vortex $k = -1$ on the top face. The $\epsilon$-tensor conspires in such a way as to have the same sign for both non-zero $k$'s, leading to, dropping the tilde from now on:

$$j_1(x) = 1, \quad j_2(x) = 0, \quad j_3(x) = -1,$$
$$j_1(x - \hat{1}) = j_2(x - \hat{2}) = j_3(x - \hat{3}) = 0. \tag{6}$$

So we see that a single loop may consist of links $j = 1$, as well as links $j = -1$. Moreover, a link $j = 1$ may be related to a vortex or an anti-vortex. If we now assign a direction to the links $j$, we see that the conservation law (4) implies that the number of links pointing to $x$ is equal to the number of links pointing away from $x$. Although for non-intersecting planar loops, we can distinguish between loops going around in a clockwise fashion from loops going around in an anti-clockwise fashion, this becomes impossible when these loops become non-planar, or intersect a number of times. Therefore, we shall make no distinction whatsoever between different loops.

In order to study the behavior of these loops, we define the following two correlation functions [3,5]:

$$C_{\mathrm{s}}(r) = \sum_{\substack{x,y,i \\ |x-y|=r}} \delta_{x \in l_i} \delta_{y \in l_i} \Big/ \sum_{\substack{x,y \\ |x-y|=r}} 1, \tag{7a}$$

$$C_{\mathrm{d}}(r) = \sum_{\substack{x,y,i,j \\ |x-y|=r}} \delta_{x \in l_i} \delta_{y \in l_j} \Big/ \sum_{\substack{x,y \\ |x-y|=r}} 1 \; - \; \Big(\frac{1}{V} \sum_{x,i} \delta_{x \in l_i}\Big)^2. \tag{7b}$$

$C_{\mathrm{s}}(r)$ yields the probability that two points $x$ and $y$, a distance $r$ away from one another, are on the same loop $l_i$. $C_{\mathrm{d}}(r)$ measures correlations between possibly different loops. $C_{\mathrm{s}}$ also has a field theoretic connection [5]. If we describe these links $j$ by a complex scalar field $\varphi$, so $j_\mu(x) = \varphi(x)\varphi^+(x+\hat{\mu})$, then we find

$$C_{\mathrm{s}}(r) = \Big\langle \varphi^+(x)\varphi(x)\varphi^+(y)\varphi(y) \Big\rangle_{\mathrm{C}} \Big|_{|x-y|=r},$$

where the subscript C stands for 'connected part'. From the fact that we can only have an even number of links at each point $x$ we can deduce that only four point and six point interactions are allowed.

Now, when we have to deal with percolating loops, i.e. loops wrapping around the lattice, we will find that $C_{\mathrm{s}}(r)$ will tend to a constant for large separations $r$. This constant is of course related to the density of these percolating loops.

However, for $C_{\mathrm{d}}(r)$, this constant is cancelled by the second term in (7b). So, in general we expect the correlation functions (7a) and (7b) to behave like

$$C(r) = d\,\frac{\exp(-ar)}{r^b} + c. \tag{7c}$$

If $C_{\mathrm{s}}$ is dominated by the first terms of the expansion of $< \varphi^+(x)\varphi(x)\varphi^+(y)\varphi(y) >$, i.e. eye-like diagrams, see figure 2a, we expect from the dimensionality of the field $\varphi$ the parameter $b$ to be equal to 2. However, if also (very) high order terms have a significant contribution to $< \varphi^+(x)\varphi(x)\varphi^+(y)\varphi(y) >$ in this expansion, see figure 2b, then it bears more resemblence to a bound state, and therefore we expect $b$ to be equal to 1.

In the spin wave phase, we expect the constant $c$ in (7c) to be equal to zero. Then, if we do not find any difference in behavior between $C_s$ and $C_d$, so in other words, if we find the same values for the parameters $a$ and $b$, then both correlation functions measure the same type of correlations – self-correlations. This then implies that the loops do not interact with one another.

In the vortex phase, we do expect different behavior when the vortex loops percolate, as then $C_s$ will tend to a constant for large separations, while this constant is absent for $C_d$.

As there is nothing to prevent loops from intersecting, we expect loops to do so. However, there is no clear way to distinguish between a self-intersecting loop and two loops touching one another. There are several ways to deal with this problem. We feel that in such a case, we should average over all different ways of dividing up such a loop. The reason is the following. If we start following a vortex loop we will arrive at a point where two or three links are pointing outwards. We then have to decide which link to follow. This choice will determine whether we follow the entire loop, or split the loop in two (or more) smaller loops. This is most easily visualized by considering the shape 8. Arriving at the crossing we have to decide in which direction to continue. One choice will lead to following the whole 8, the other will lead to splitting the shape up into two small rings.

Having defined loops, we can also determine the distribution of loop lengths $\rho(l)$. Again, when loops percolate, we expect loops of any length in the system, hence the distribution $\rho(l)$ will tend to a (non-zero) constant for large lengths $l$.

## 3 Results

We used a 10 hit Metropolis algorithm with site dependent step size, combined with $\gamma = 2$ multigrid, as described in detail in [7]. We made runs consisting of 10 batches of 100 measurements each, after discarding 500 measurements for thermalization. The measurements were taken after each time the spins on the original lattice were updated, i.e., after each half W-cycle. Both $C_s$ and $C_d$ were obtained from averaging over the 10 batches.

We made runs for $\beta = 0.00$ up to $\beta = 1.00$ in steps of $\Delta\beta = 0.05$, as well as runs with $\Delta\beta = 0.01$ for $\beta$ between 0.40 and 0.50. Two additional runs were made at $\beta = 0.525$ and at $\beta = 0.575$. The lattice sizes we used were $6^3$, $8^3$, $12^3$, and $16^3$.

To define a vortex loop, we choose the following strategy. We start at some point $x_0$, which happens to have a link $j \neq 0$. This point $x_0$ will obviously be on a vortex loop. We then scan in a prefixed order for the next point of the vortex loop, marking the connecting link as read. This search is repeated until no more unread links are to be found. We have then returned to the starting point $x_0$. We then search for a next point $x_1$, and repeat the procedure. If we visit some point $x'$, to which four (or six) non-zero links are connected, there is the possibility that this self-intersecting loop will be split into two (or more) different loops, depending on the geometry of the original, self-intersecting loop. Since the system is isotropic, there is no preferred geometry, and this procedure will in effect average over the different possibilities of splitting up loops (including no splits at all).

Apart from the correlation functions $C_s$ and $C_d$ and the length distribution $\rho(l)$, we also measured the average number of loops $\#$, the average loop length $\Lambda$, and the fractal dimension $d_f$. The latter is defined as the ratio of the total number of links over the total number of points visited by the vortex loops [5].

### 3.1 $C_s$ and $C_d$

In figure 3, we show the behavior of $C_s$ on a $16^3$ lattice, For $\beta$ in the range $\beta = 0.40 - 0.50$. As can be seen from this figure, there is a difference in behavior between $\beta < \beta_c$ and $\beta > \beta_c$ ($\beta_c \approx 0.4542$ [8–10]). Notice that $C_s$ approaches a constant for $\beta \lesssim \beta_c$, but that for $\beta > \beta_c$, this constant is absent. When making a $\chi^2$-fit of $C_s(r)$ to (7c), we find the results as given in tables 1 to 4. It is interesting to note that for $\beta < \beta_c$ the parameters $a$, $b$ and $d$ are relatively independent of $\beta$, while only $c$ does depend on $\beta$. As a matter of fact, $c$ is monotonically decreasing for increasing $\beta$. This means, since $c$ is related to the density of percolating loops, that also this density is decreasing for increasing $\beta$, and vanishes when $c$ vanishes.

Figures 4a and 4b show $C_s$ and $C_d$ for a $16^3$ lattice at $\beta = 0.30$ and $\beta = 0.55$, respectively. It is observed that $C_d$ falls off exponentially on either side of the phase transition, and that for $\beta > \beta_c$, both $C_s$ and $C_d$ fall off with the same exponent. We did not fit $C_d$ to (7c) since there were not enough data points to make a reasonable fit.

In figure 5, we have shown the exponent $a$ versus $\beta$ in the range $\beta = 0.35$ to $\beta = 0.60$, for all four lattice sizes. It can be seen from this figure that $a$ starts to rise approximately linearly for $\beta \geq \beta_c$, up to $\beta \approx 0.60$. Figure 6a shows the constant $c$ versus $\beta$ for $\beta = 0.00$ to $\beta = 0.60$, while 6b shows $c$ in the range $\beta = 0.35$ to $\beta = 0.50$. We see that the results for both $a$ and $c$ for the lattice sizes $N = 12$ and $N = 16$ are within one another's error bars.

The upshot of these results is that there is no clear sign for interaction between loops on either side of the phase transition. In the spin wave phase, we see that $C_s(r)$ behaves identical to $C_d(r)$, while in the vortex phase $C_d(r)$ falls off with a large exponent. An estimate from figure 4a gives $C_d(r) \approx \exp(-15\,r)$.

Focussing on $C_s$, we see that the mass $a$ vanishes in the vortex phase. From the fact that $b \approx 1.5$, we can deduce that the field $\varphi$ does form a bound state, albeit a weakly bound one. Finding a non-zero value for $c$ in the vortex phase, and a zero value in the spin wave phase means that the system undergoes symmetry breaking, and the non-zero value of $c$ indicates that a condensate has been formed.

The fact that the results for $N = 12$ are equal to the ones from $N = 16$ implies that we do not need to study larger lattices if we are only interested in the behavior of vortex loops.

## 3.2 $\#$, $\Lambda$, and $d_f$.

In tables 5 and 6, we give the results for $\#$, $\Lambda$, and $d_f$. The errors were obtained by blocking. This information is also displayed in figures 7a–d. Several interesting observations can be made. The first one is that the density of loops, $\Omega = \#/N^3$, falls off exponentially for $\beta \gg \beta_c$. This means that loops become exponentially suppressed in this coupling region. This is hardly surprising as this region is the spin wave phase, where the dominant degrees of freedom are spin waves. Note that here $\Omega$ does not show much finite size effect, in contrast to the vortex phase, $\beta \ll \beta_c$, where indeed there are finite size effects. The second observation is the similarity between $\Lambda$ and $d_f$. It appears that, on average, the larger the loops, the more they (self)intersect. Again note that the results for $N = 12$ and $N = 16$ are at most a few $\sigma$ away from one another in the coupling region $0.30 < \beta < 0.60$, supporting the observation made earlier that these lattice sizes are large enough to obtain reliable results for vortex loops.

## 3.3 $\rho(l)$ and $\bar{l}$.

Figures 8a and 8b depict the behavior of $\rho(l)$ versus $l$ for $16^3$ lattices at $\beta = 0.30$ and $\beta = 0.60$, respectively. Again we see different behavior on either side of the phase transition. The presence of a constant for $\beta < \beta_c$ is clear, while for $\beta \gg \beta_c$, there is an exponential suppression of larger loops. Notice that there still are a small number of loops with lengths a few times the linear extension of the lattice. In tables 7 and 8, we present the results for fitting $\rho(l)$ to (7c). For $16^3$ we have plotted $a$ and $b$ from this fit in figure 9. Note that both $a$ and $b$ behave (roughly) linearly for $0.45 \leq \beta \leq 0.60$. An estimate would give $a \approx 2.36\beta - 1.14$, and $b \approx 6.8 - 9.8\beta$. The fact that the parameter $a$ becomes non-zero for $\beta > \beta_c$ is, of course, related to the exponential suppression of vortex loops found in the previous section.

Turning our attention now to the vortex phase, we see that the parameter $b$ is constant, albeit with large errors. An estimate of $b$ in this region would give $b \approx 2.2(3)$. For $a$, we would estimate $a \approx 0.01(7)$. These values are consistent with the results from [11]. The results for the spin wave phase do not agree, mainly due to the absence of an exponential factor in [11].

As we have made 10 batches per run, we can average over the length of the largest loop per batch to get an estimate for the length of the largest loop for a given $\beta$-value. This average length, $\overline{\ell}$, is presented in table 9. We have also made a log-plot of these data in figure 10. It should be kept in mind when reading tables 7 and 8 that $l$ runs essentially from 0 to $\overline{\ell}$. Notice that $\overline{\ell}$ scales differently on either side of the phase transition.

# 4 Conclusions and discussions

The picture of the phase transition is now quite clear. As $\beta \searrow \beta_c$, we see that the loop density $\Omega$ increases, while the average length $\Lambda$ remains constant, or increases slightly. As $\beta > \beta_c$, but quite close to $\beta_c$, the loop density remains constant, while the average length increases sharply. If we make an estimate in this region of how many loops of length $\Lambda$ can fit in, say, a $16^3$ lattice, we find that at $\beta = 0.48$ there are on average 125 loops in the system. The average length at that $\beta$-value is 10.6. We can accommodate a loop of such a length in a $3^3$ block. So we see that about the maximum number of loops are present in the system. As $\beta$ is further decreased into the vortex regime, we see that the loop density remains roughly constant, while the average length increases roughly linearly with decreasing $\beta$. So around the phase transition, the length of the loops is becoming so large that different loops start touching one another, and will fuse to form one, large loop. This loop will necessarily have many self-intersections, which can be witnessed from the behavior of $d_f$ in this region, and from the similarity between $\Lambda$ and $d_f$. At the phase transition, finally, these long loops start percolating through the lattice. This is substantiated by the fact that for $\beta \leq \beta_c$, $C_s(r) \to c \neq 0$ for large $r$.

All this also indicates that the different loops interact rather feebly with one another, if at all. If they would have interacted more strongly with one another, we would expect only a handful of larger loops at $\beta$-values slightly above $\beta_c$, instead of roughly 100 short loops, filling the whole lattice. Another clue is the fact that $C_s$ and $C_d$ fall off with the same exponent for $\beta > \beta_c$. Figure 4b shows both at $\beta = 0.55$. For this $\beta$-value, there are on average 80 loops in the system. Had the loops interacted with one another, we would have seen a difference in behavior between the two correlation functions. This is, of course, not very surprising. Since a single loop may consist of links $j = 1$, as well as of links $j = -1$, we expect no significant interaction between links of opposite sign, nor of like sign. Any interactions between links would lead to the collapse of the loops they constitute.

The onset of percolation also explains the absence of an exponential in $C_s$ and $\rho$ for $\beta < \beta_c$. As percolating loops destroy any long range order, these exponentials cannot be accustomed for, and are absent, or small, due to finite size effects.

The description in terms of the field $\varphi$ is the following. For couplings above $\beta_c$, the system exhibits a finite mass $a$. At $\beta_c$, the system undergoes a symmetry breaking phase transition, at which the mass vanishes. Furthermore, a condensate is formed, which is witnessed by the constant $c$. In this condensate, the $\varphi$ fields form weakly bound states.

So the picture of the phase transition in the 3-d XY model is the following. At large couplings $\beta$, the system is dominated by spin waves, and allows only a few, small vortex loops. As the coupling $\beta$ is lowered, the number of loops increases, but the average length remains constant. As $\beta$ approaches $\beta_c$, the number of loops becomes so large that they start to fuse, and form longer, self-intersecting loops. At $\beta_c$, these loops start to percolate through the lattice. For $\beta$ below $\beta_c$, the system is described by loops of all sizes up to $N^3$. For all couplings $\beta$, there is no noticeable interaction between the vortex loops.

The correlation functions $C_s$ and $C_d$ can be extended to other objects which are invariant under the symmetries of the system. More specific, it would be possible to extend $C_s$ and $C_d$ to measure the behavior of monopoles in three (or more) dimensional U(1) gauge theory. The field theoretic connection of $C_s$ would then allow us to study the behavior of monopoles in these theories, hopefully including the mechanism of monopole condensation. The more obvious use of $C_s$ and $C_d$ for Dirac strings will fail, as these strings are gauge dependent.

# ACKNOWLEDGEMENTS

We would like to thank Chris Michael, Jan Smit, Ronald Rietman and Peer Ueberholz for useful discussions. This work was supported by EC contract SC1 *CT91-0642.

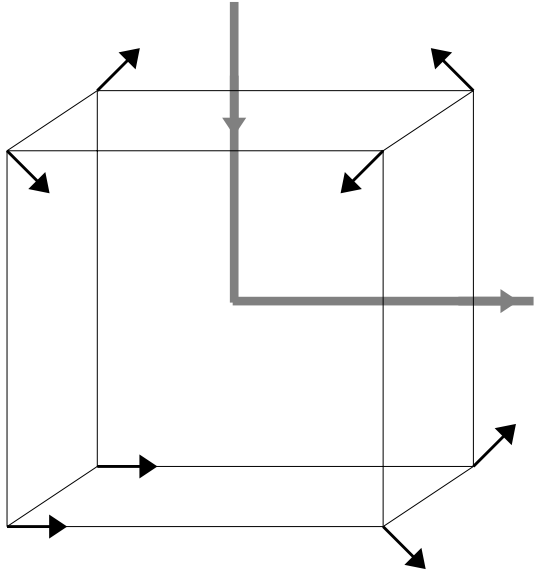

Figure 1: Spin configuration as given in (5). The spins are in the 1-3 plane. The links $j$ following from these spins are shown in gray. A link $j_\mu(x) = 1$ is chosen to point away from $x$.

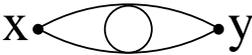

Fig. 2a

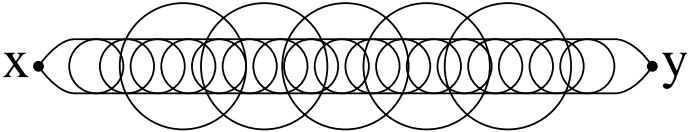

Fig. 2b

Figure 2: Possible diagrams that constitute $< \varphi^+(x)\varphi(x)\varphi^+(y)\varphi(y) >$. Upper, figure 2a, a diagram leading to $b = 2$ in equation 7c. Lower, figure 2b, a diagram that describes meson-like behavior of the aforementioned correlation function, and would lead to $b = 1$.

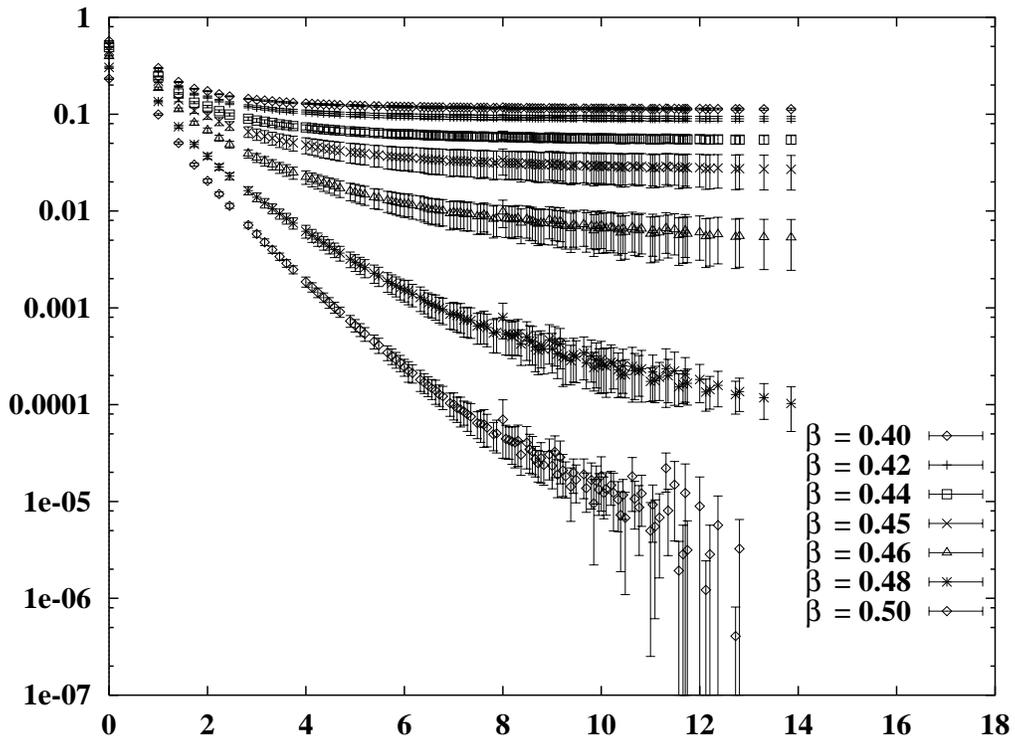

Figure 3: The correlation function $C_S(r)$ versus $r$, from a $16^3$ lattice. The $\beta$-values are shown.

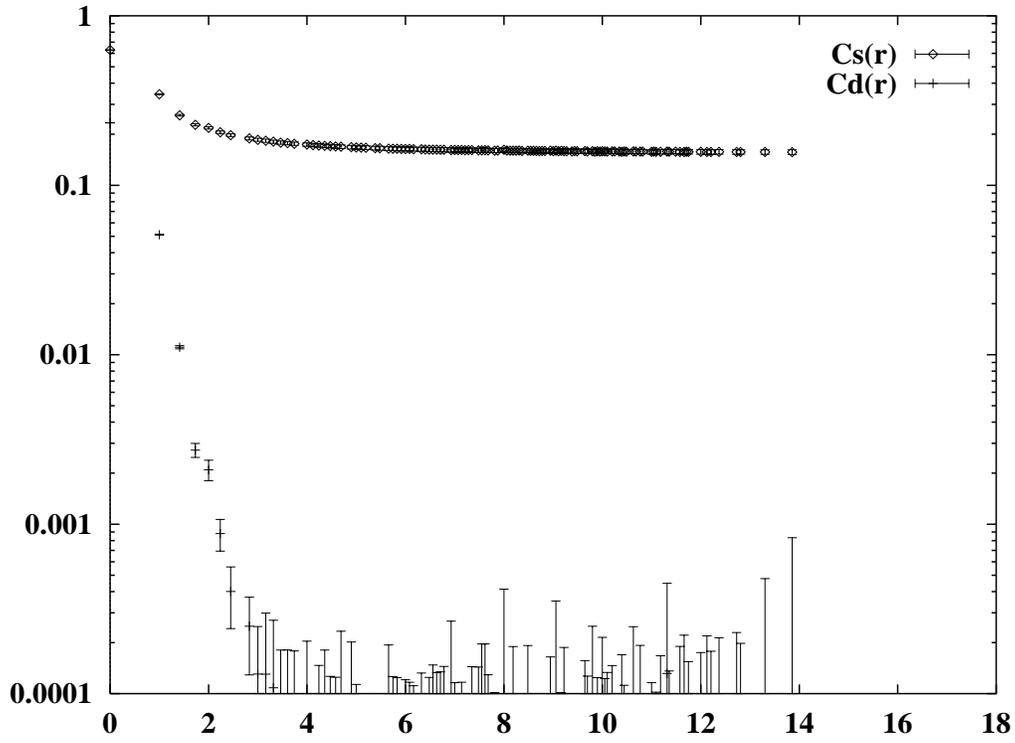

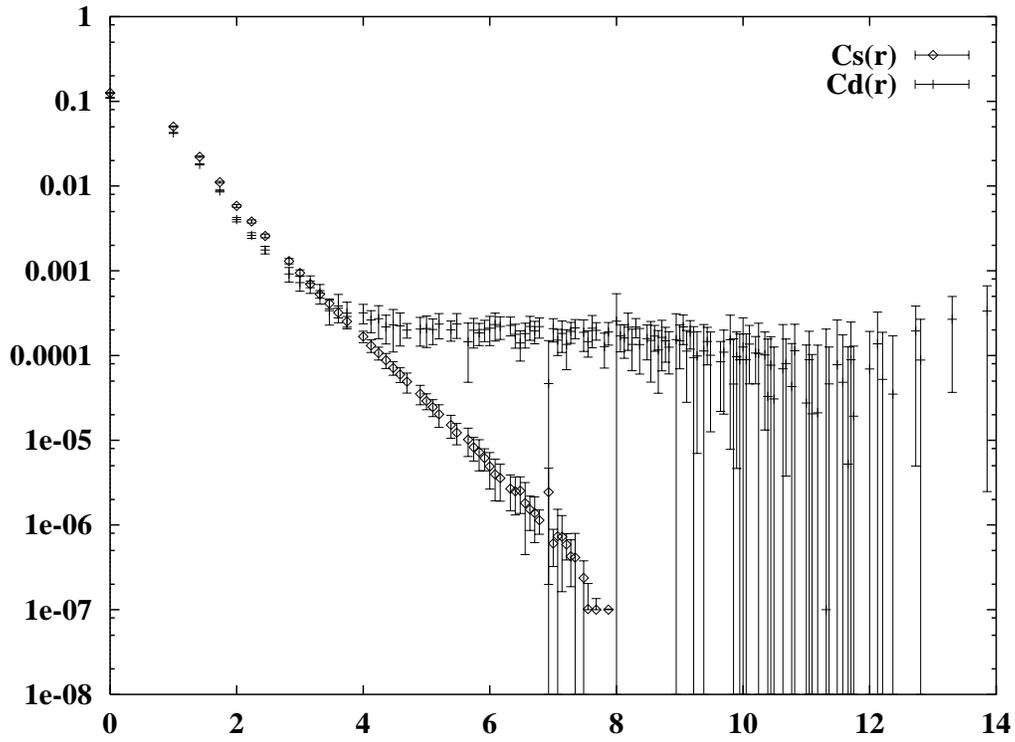

Figure 4: The correlation functions $C_s(r)$ and $C_d$ versus $r$, from a $16^3$ lattice for $\beta = 0.30$ (figure 4a, upper), and for $\beta = 0.55$ (figure 4b, lower).

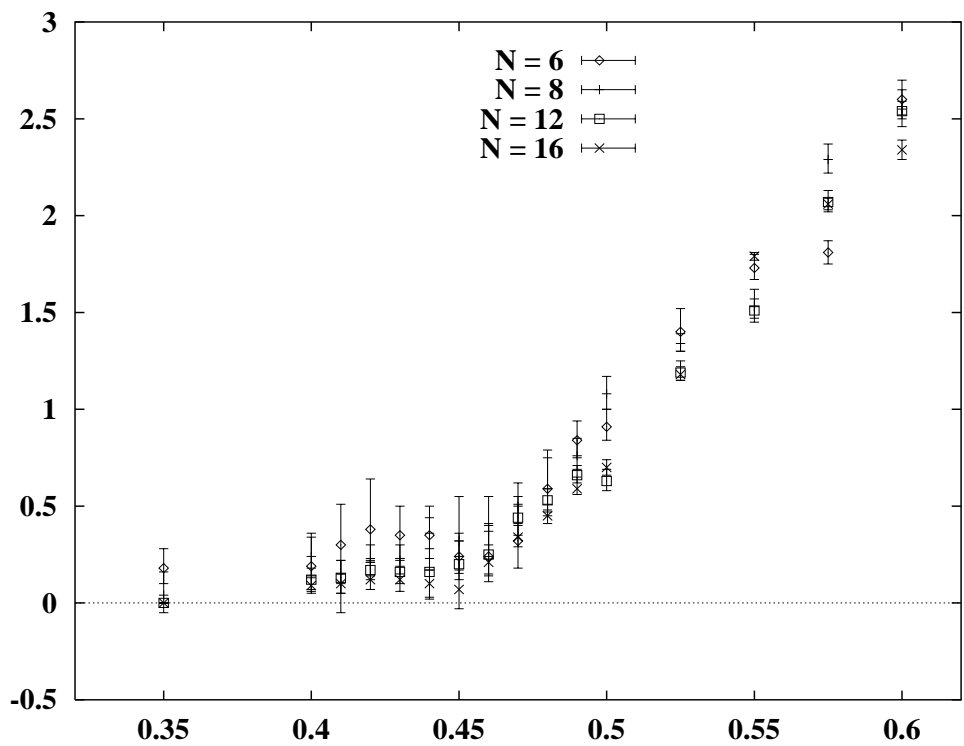

Figure 5: The exponent $a$ from different lattice sizes, depicted versus $\beta$.

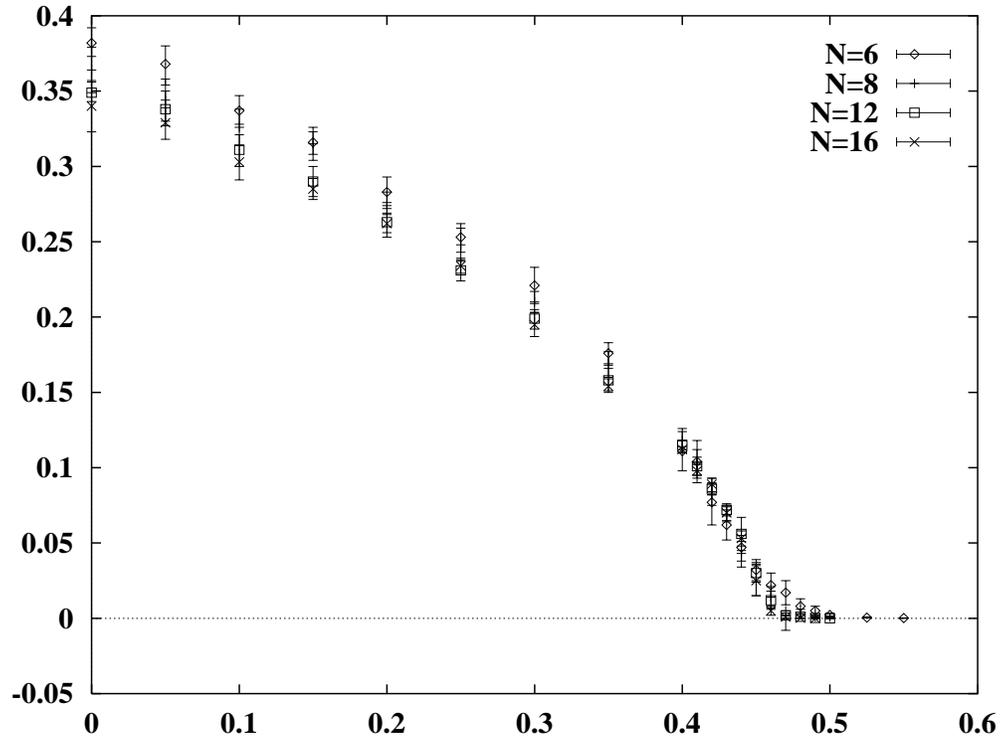

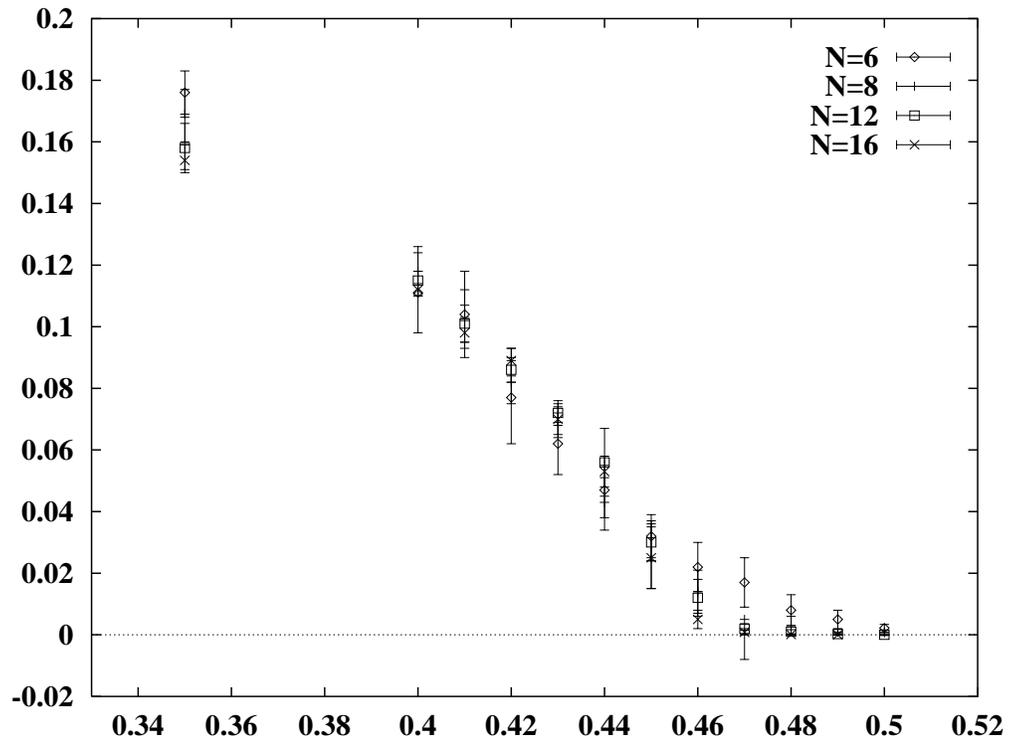

Figure 6: The constant $c$ from different lattice sizes, depicted versus $\beta$. in the range $\beta = 0.00$ to $\beta = 0.60$ (figure 6a, upper), and for $\beta$ in the range $\beta = 0.35$ to $\beta = 0.60$ (figure 6b, lower).

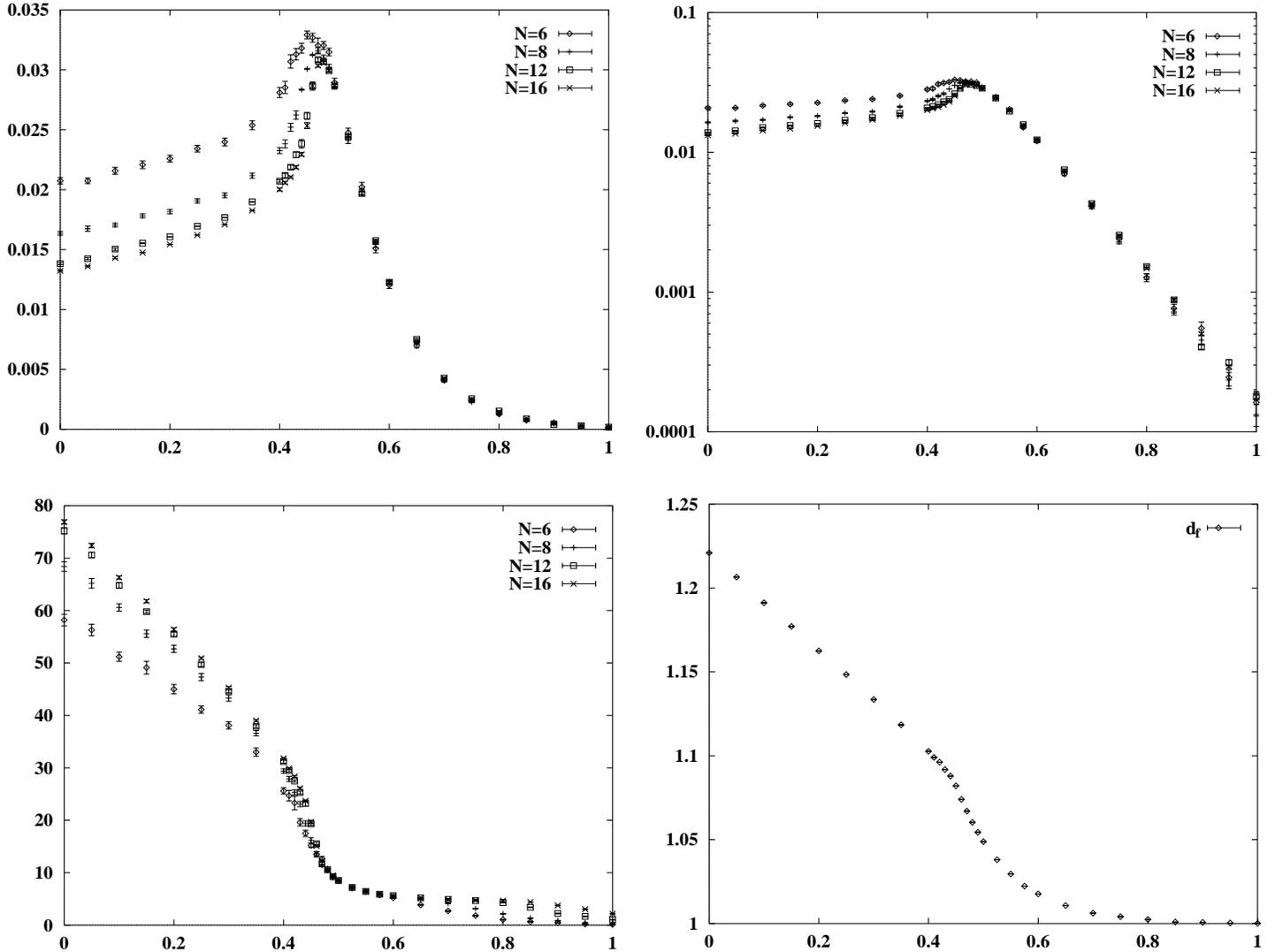

Figure 7: The results for the loop density $\Omega = \#/N^3$ on a linear scale (top left, figure 7a) and on a logarithmic scale (top right, figure 7b), and the average loop length $\Lambda$ (bottom left, figure 7c) from all four lattice sizes, and the fractal dimension $d_\mathrm{f}$ from $16^3$ only (bottom right, figure 7d). All quantities are shown versus $\beta$.

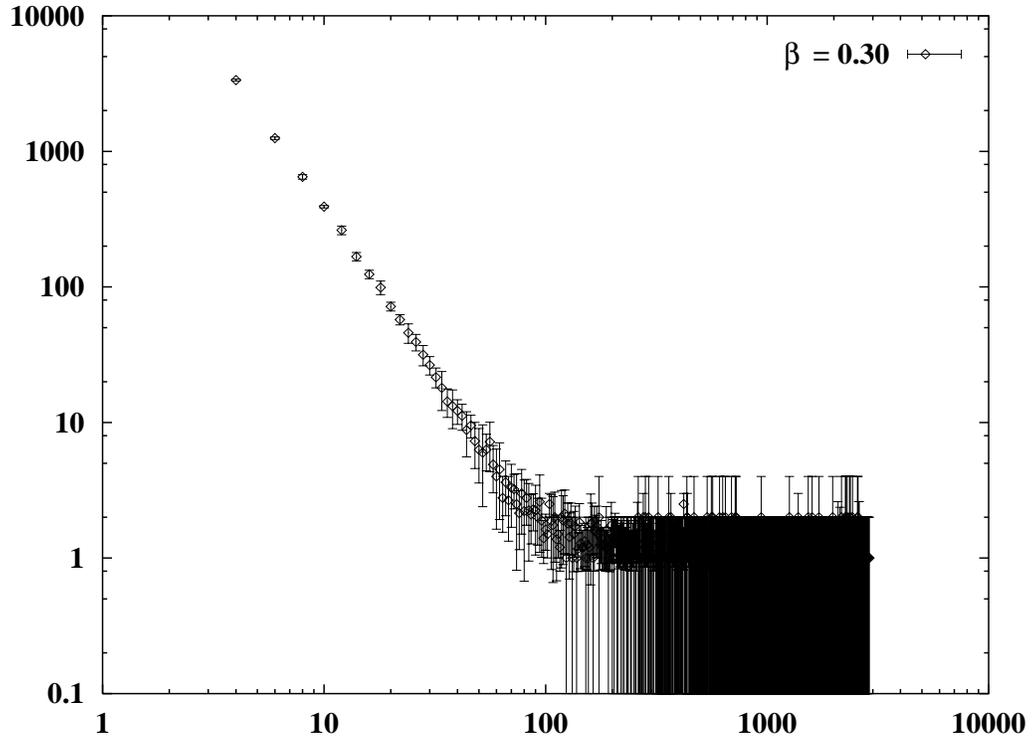
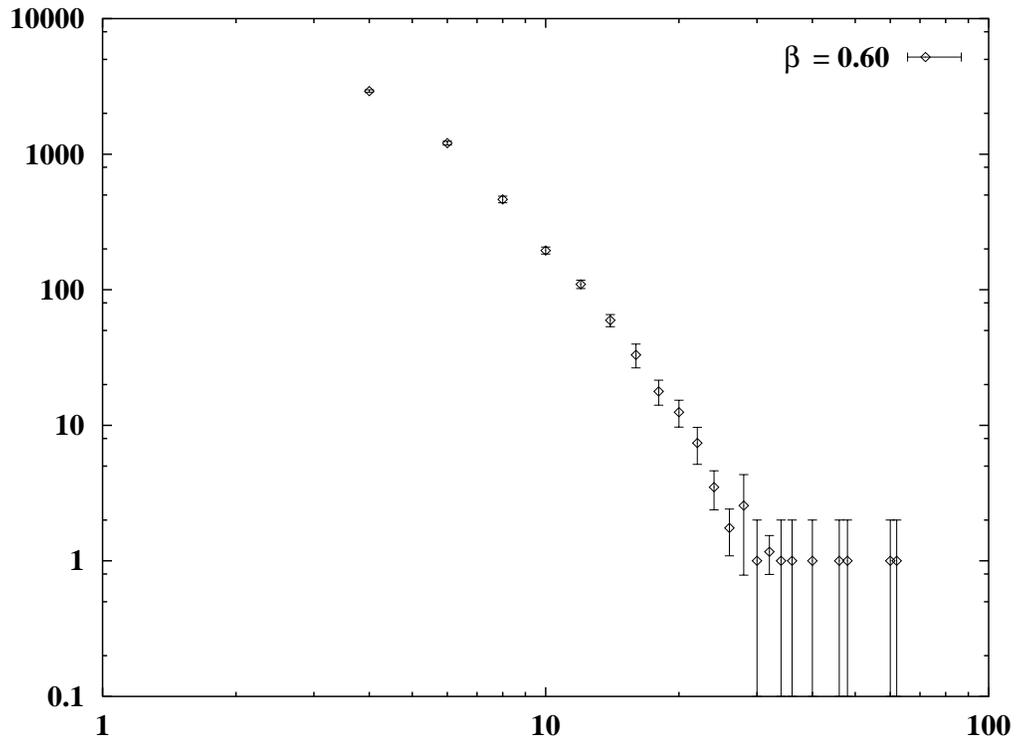

Figure 8: The loop density $\rho(l)$ versus $l$ from $16^3$ lattices for $\beta = 0.30$ (upper, figure 8a), and for $\beta = 0.60$ (lower, figure 8b).

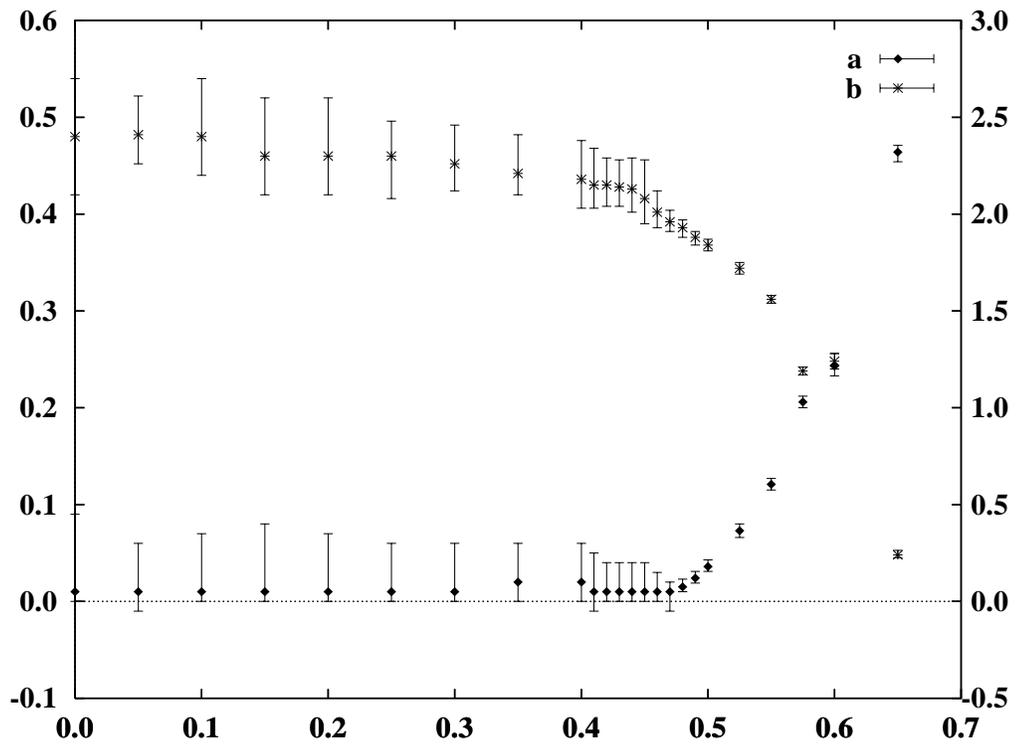

Figure 9: The behavior of the exponent $a$ (left scale), and the power $b$ (right scale) from fitting $\rho(l)$ to (7c), from $16^3$. Both parameters are plotted versus $\beta$.

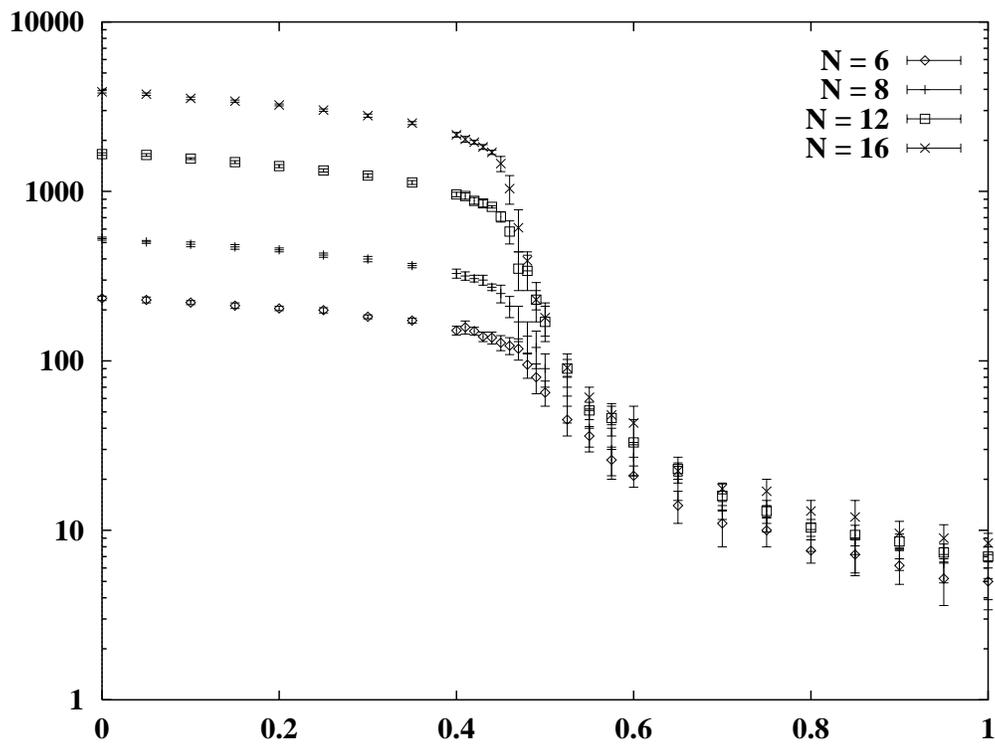

Figure 10: The average length of the largest loop, $\overline{\ell}$, versus $\beta$ on a log-plot, from all four lattice sizes.

| $\beta$ | $a$ | $b$ | $c$ | $d$ |
|---|---|---|---|---|
| 0.00 | — | $1.9^{+0.6}_{-0.4}$ | $0.382^{+0.010}_{-0.009}$ | $0.15^{+0.02}_{-0.02}$ |
| 0.05 | — | $1.9^{+1.0}_{-0.5}$ | $0.368^{+0.012}_{-0.014}$ | $0.15^{+0.03}_{-0.03}$ |
| 0.10 | — | $2.0^{+0.6}_{-0.4}$ | $0.337^{+0.010}_{-0.009}$ | $0.15^{+0.03}_{-0.02}$ |
| 0.15 | — | $2.0^{+0.5}_{-0.3}$ | $0.316^{+0.007}_{-0.008}$ | $0.16^{+0.02}_{-0.02}$ |
| 0.20 | — | $2.0^{+0.7}_{-0.5}$ | $0.283^{+0.010}_{-0.011}$ | $0.16^{+0.03}_{-0.03}$ |
| 0.25 | — | $2.0^{+0.5}_{-0.4}$ | $0.253^{+0.009}_{-0.010}$ | $0.16^{+0.02}_{-0.02}$ |
| 0.30 | — | $2.0^{+0.8}_{-0.5}$ | $0.221^{+0.012}_{-0.012}$ | $0.16^{+0.04}_{-0.03}$ |
| 0.35 | $0.18^{+0.10}_{-0.08}$ | $1.7^{+0.5}_{-0.3}$ | $0.176^{+0.007}_{-0.008}$ | $0.20^{+0.02}_{-0.02}$ |
| 0.40 | $0.19^{+0.17}_{-0.13}$ | $1.6^{+0.8}_{-0.5}$ | $0.111^{+0.013}_{-0.013}$ | $0.21^{+0.05}_{-0.04}$ |
| 0.41 | $0.30^{+0.21}_{-0.15}$ | $1.5^{+1.0}_{-0.5}$ | $0.104^{+0.014}_{-0.014}$ | $0.23^{+0.06}_{-0.06}$ |
| 0.42 | $0.38^{+0.26}_{-0.17}$ | $1.4^{+1.2}_{-0.6}$ | $0.077^{+0.016}_{-0.015}$ | $0.25^{+0.07}_{-0.07}$ |
| 0.43 | $0.35^{+0.15}_{-0.12}$ | $1.4^{+0.6}_{-0.4}$ | $0.062^{+0.010}_{-0.010}$ | $0.24^{+0.05}_{-0.05}$ |
| 0.44 | $0.35^{+0.15}_{-0.12}$ | $1.4^{+0.6}_{-0.4}$ | $0.047^{+0.008}_{-0.009}$ | $0.23^{+0.05}_{-0.04}$ |
| 0.45 | $0.24^{+0.12}_{-0.09}$ | $1.5^{+0.5}_{-0.3}$ | $0.032^{+0.007}_{-0.008}$ | $0.20^{+0.03}_{-0.03}$ |
| 0.46 | $0.24^{+0.17}_{-0.13}$ | $1.6^{+0.6}_{-0.4}$ | $0.022^{+0.008}_{-0.008}$ | $0.19^{+0.04}_{-0.05}$ |
| 0.47 | $0.32^{+0.19}_{-0.14}$ | $1.6^{+0.7}_{-0.4}$ | $0.017^{+0.008}_{-0.008}$ | $0.19^{+0.05}_{-0.05}$ |
| 0.48 | $0.59^{+0.20}_{-0.15}$ | $1.3^{+0.7}_{-0.5}$ | $0.008^{+0.005}_{-0.005}$ | $0.22^{+0.06}_{-0.06}$ |
| 0.49 | $0.84^{+0.10}_{-0.09}$ | $1.1^{+0.3}_{-0.3}$ | $0.005^{+0.003}_{-0.003}$ | $0.25^{+0.04}_{-0.02}$ |
| 0.50 | $0.91^{+0.09}_{-0.07}$ | $1.1^{+0.3}_{-0.2}$ | $0.0022^{+0.0012}_{-0.0013}$ | $0.23^{+0.03}_{-0.03}$ |
| 0.525 | $1.40^{+0.12}_{-0.10}$ | $0.8^{+0.4}_{-0.3}$ | $0.0005^{+0.0004}_{-0.0003}$ | $0.28^{+0.05}_{-0.05}$ |
| 0.55 | $1.73^{+0.08}_{-0.06}$ | $0.5^{+0.3}_{-0.2}$ | $0.00015^{+0.00015}_{-0.00016}$ | $0.29^{+0.03}_{-0.03}$ |
| 0.575 | $1.81^{+0.06}_{-0.06}$ | $0.39^{+0.11}_{-0.11}$ | — | $0.288^{+0.019}_{-0.018}$ |
| 0.60 | $2.60^{+0.10}_{-0.08}$ | $0.0^{+0.3}_{-0.3}$ | — | $0.35^{+0.04}_{-0.05}$ |
| 0.65 | $3.18^{+0.10}_{-0.08}$ | $0.0^{+0.3}_{-0.3}$ | — | $0.34^{+0.04}_{-0.05}$ |
| 0.70 | $4.12^{+0.16}_{-0.12}$ | — | — | $0.50^{+0.09}_{-0.09}$ |
| 0.75 | $3.92^{+0.16}_{-0.13}$ | — | — | $0.23^{+0.04}_{-0.05}$ |

Table 1: The results from fitting $C_S$ to (7c) for $6^3$.

| $\beta$ | $a$ | $b$ | $c$ | $d$ |
|---|---|---|---|---|
| 0.00 | — | $1.8^{+1.1}_{-0.6}$ | $0.364^{+0.015}_{-0.015}$ | $0.16^{+0.05}_{-0.05}$ |
| 0.05 | — | $1.7^{+0.5}_{-0.4}$ | $0.344^{+0.014}_{-0.014}$ | $0.17^{+0.03}_{-0.03}$ |
| 0.10 | — | $1.8^{+0.9}_{-0.4}$ | $0.326^{+0.012}_{-0.012}$ | $0.16^{+0.03}_{-0.03}$ |
| 0.15 | — | $1.9^{+0.9}_{-0.5}$ | $0.315^{+0.011}_{-0.011}$ | $0.16^{+0.04}_{-0.04}$ |
| 0.20 | — | $1.8^{+0.5}_{-0.3}$ | $0.276^{+0.007}_{-0.008}$ | $0.17^{+0.02}_{-0.02}$ |
| 0.25 | — | $1.9^{+0.7}_{-0.4}$ | $0.248^{+0.011}_{-0.011}$ | $0.17^{+0.03}_{-0.03}$ |
| 0.30 | — | $1.8^{+0.4}_{-0.3}$ | $0.210^{+0.007}_{-0.007}$ | $0.17^{+0.02}_{-0.02}$ |
| 0.35 | $0.04^{+0.12}_{-0.09}$ | $1.8^{+0.5}_{-0.3}$ | $0.169^{+0.008}_{-0.009}$ | $0.19^{+0.03}_{-0.03}$ |
| 0.40 | $0.24^{+0.10}_{-0.09}$ | $1.5^{+0.5}_{-0.3}$ | $0.118^{+0.008}_{-0.008}$ | $0.23^{+0.02}_{-0.03}$ |
| 0.41 | $0.10^{+0.12}_{-0.15}$ | $1.7^{+0.5}_{-0.4}$ | $0.103^{+0.009}_{-0.010}$ | $0.20^{+0.03}_{-0.03}$ |
| 0.42 | $0.21^{+0.09}_{-0.07}$ | $1.5^{+0.3}_{-0.3}$ | $0.082^{+0.007}_{-0.007}$ | $0.23^{+0.03}_{-0.03}$ |
| 0.43 | $0.22^{+0.08}_{-0.07}$ | $1.4^{+0.3}_{-0.2}$ | $0.069^{+0.005}_{-0.005}$ | $0.23^{+0.02}_{-0.03}$ |
| 0.44 | $0.28^{+0.16}_{-0.11}$ | $1.3^{+0.6}_{-0.3}$ | $0.043^{+0.008}_{-0.009}$ | $0.24^{+0.05}_{-0.06}$ |
| 0.45 | $0.32^{+0.23}_{-0.15}$ | $1.3^{+0.9}_{-0.5}$ | $0.025^{+0.011}_{-0.010}$ | $0.24^{+0.07}_{-0.08}$ |
| 0.46 | $0.37^{+0.18}_{-0.13}$ | $1.3^{+0.7}_{-0.4}$ | $0.014^{+0.007}_{-0.007}$ | $0.23^{+0.05}_{-0.07}$ |
| 0.47 | $0.50^{+0.12}_{-0.09}$ | $1.2^{+0.4}_{-0.2}$ | $0.005^{+0.003}_{-0.003}$ | $0.23^{+0.05}_{-0.04}$ |
| 0.48 | $0.59^{+0.16}_{-0.12}$ | $1.2^{+0.5}_{-0.4}$ | $0.003^{+0.003}_{-0.003}$ | $0.23^{+0.06}_{-0.05}$ |
| 0.49 | $0.76^{+0.09}_{-0.07}$ | $1.1^{+0.3}_{-0.2}$ | $0.0009^{+0.0009}_{-0.0008}$ | $0.23^{+0.04}_{-0.03}$ |
| 0.50 | $1.08^{+0.09}_{-0.08}$ | $0.8^{+0.4}_{-0.3}$ | $0.0007^{+0.0009}_{-0.0009}$ | $0.28^{+0.03}_{-0.04}$ |
| 0.525 | $1.34^{+0.05}_{-0.04}$ | $0.85^{+0.15}_{-0.14}$ | — | $0.254^{+0.018}_{-0.018}$ |
| 0.55 | $1.53^{+0.09}_{-0.06}$ | $0.9^{+0.3}_{-0.2}$ | — | $0.23^{+0.03}_{-0.02}$ |
| 0.575 | $2.29^{+0.08}_{-0.07}$ | $0.2^{+0.2}_{-0.2}$ | — | $0.35^{+0.04}_{-0.04}$ |
| 0.60 | $2.55^{+0.10}_{-0.09}$ | — | — | $0.35^{+0.05}_{-0.05}$ |
| 0.65 | $3.04^{+0.07}_{-0.07}$ | — | — | $0.30^{+0.03}_{-0.03}$ |
| 0.70 | $3.30^{+0.07}_{-0.07}$ | — | — | $0.200^{+0.017}_{-0.017}$ |
| 0.75 | $3.80^{+0.10}_{-0.09}$ | — | — | $0.19^{+0.02}_{-0.02}$ |
| 0.80 | $4.00^{+0.09}_{-0.08}$ | — | — | $0.128^{+0.014}_{-0.013}$ |

Table 2: The results from fitting $C_{\mathrm{S}}$ to (7c) for $8^3$.

| $\beta$ | $a$ | $b$ | $c$ | $d$ |
|---|---|---|---|---|
| 0.00 | — | $1.6^{+0.4}_{-0.3}$ | $0.349^{+0.007}_{-0.006}$ | $0.18^{+0.02}_{-0.03}$ |
| 0.05 | — | $1.7^{+0.7}_{-0.4}$ | $0.338^{+0.012}_{-0.011}$ | $0.17^{+0.04}_{-0.04}$ |
| 0.10 | — | $1.7^{+0.7}_{-0.4}$ | $0.311^{+0.010}_{-0.011}$ | $0.18^{+0.04}_{-0.03}$ |
| 0.15 | — | $1.7^{+0.7}_{-0.4}$ | $0.290^{+0.010}_{-0.010}$ | $0.18^{+0.03}_{-0.04}$ |
| 0.20 | — | $1.7^{+0.7}_{-0.3}$ | $0.263^{+0.011}_{-0.010}$ | $0.18^{+0.04}_{-0.04}$ |
| 0.25 | — | $1.7^{+0.4}_{-0.3}$ | $0.231^{+0.007}_{-0.007}$ | $0.18^{+0.03}_{-0.02}$ |
| 0.30 | — | $1.7^{+0.5}_{-0.2}$ | $0.199^{+0.006}_{-0.007}$ | $0.18^{+0.03}_{-0.02}$ |
| 0.35 | — | $1.7^{+0.5}_{-0.3}$ | $0.158^{+0.008}_{-0.007}$ | $0.18^{+0.03}_{-0.02}$ |
| 0.40 | $0.12^{+0.06}_{-0.05}$ | $1.6^{+0.2}_{-0.2}$ | $0.115^{+0.003}_{-0.005}$ | $0.21^{+0.02}_{-0.02}$ |
| 0.41 | $0.13^{+0.09}_{-0.08}$ | $1.5^{+0.4}_{-0.2}$ | $0.101^{+0.006}_{-0.006}$ | $0.21^{+0.03}_{-0.03}$ |
| 0.42 | $0.17^{+0.06}_{-0.05}$ | $1.5^{+0.2}_{-0.2}$ | $0.086^{+0.004}_{-0.004}$ | $0.22^{+0.02}_{-0.02}$ |
| 0.43 | $0.16^{+0.07}_{-0.06}$ | $1.4^{+0.3}_{-0.2}$ | $0.072^{+0.004}_{-0.004}$ | $0.22^{+0.02}_{-0.03}$ |
| 0.44 | $0.16^{+0.20}_{-0.14}$ | $1.4^{+0.8}_{-0.4}$ | $0.056^{+0.011}_{-0.011}$ | $0.22^{+0.06}_{-0.06}$ |
| 0.45 | $0.20^{+0.12}_{-0.08}$ | $1.3^{+0.4}_{-0.3}$ | $0.030^{+0.007}_{-0.006}$ | $0.23^{+0.04}_{-0.04}$ |
| 0.46 | $0.25^{+0.15}_{-0.10}$ | $1.3^{+0.4}_{-0.4}$ | $0.012^{+0.006}_{-0.006}$ | $0.22^{+0.06}_{-0.05}$ |
| 0.47 | $0.44^{+0.11}_{-0.09}$ | $1.2^{+0.4}_{-0.3}$ | $0.0019^{+0.0017}_{-0.0018}$ | $0.23^{+0.05}_{-0.05}$ |
| 0.48 | $0.53^{+0.06}_{-0.05}$ | $1.2^{+0.2}_{-0.2}$ | $0.0010^{+0.0007}_{-0.0007}$ | $0.23^{+0.02}_{-0.03}$ |
| 0.49 | $0.66^{+0.05}_{-0.04}$ | $1.17^{+0.14}_{-0.17}$ | $0.00024^{+0.00020}_{-0.00020}$ | $0.220^{+0.019}_{-0.019}$ |
| 0.50 | $0.63^{+0.06}_{-0.05}$ | $1.48^{+0.20}_{-0.16}$ | $0.00001^{+0.00005}_{-0.00004}$ | $0.19^{+0.03}_{-0.03}$ |
| 0.525 | $1.19^{+0.06}_{-0.04}$ | $0.96^{+0.17}_{-0.14}$ | — | $0.23^{+0.03}_{-0.03}$ |
| 0.55 | $1.51^{+0.06}_{-0.06}$ | $0.91^{+0.17}_{-0.15}$ | — | $0.23^{+0.02}_{-0.02}$ |
| 0.575 | $2.07^{+0.06}_{-0.05}$ | $0.37^{+0.19}_{-0.16}$ | — | $0.29^{+0.03}_{-0.03}$ |
| 0.60 | $2.54^{+0.05}_{-0.04}$ | $0.00^{+0.17}_{-0.17}$ | — | $0.34^{+0.03}_{-0.02}$ |
| 0.65 | $2.96^{+0.04}_{-0.05}$ | — | — | $0.31^{+0.02}_{-0.02}$ |
| 0.70 | $3.43^{+0.08}_{-0.07}$ | — | — | $0.26^{+0.03}_{-0.03}$ |
| 0.75 | $3.68^{+0.04}_{-0.04}$ | — | — | $0.204^{+0.010}_{-0.009}$ |
| 0.80 | $4.16^{+0.06}_{-0.07}$ | — | — | $0.180^{+0.015}_{-0.014}$ |

Table 3: The results from fitting $C_{\mathrm{S}}$ to (7c) for $12^3$.

| $\beta$ | $a$ | $b$ | $c$ | $d$ |
|---|---|---|---|---|
| 0.00 | — | $1.5^{+1.2}_{-0.4}$ | $0.340^{+0.017}_{-0.017}$ | $0.18^{+0.07}_{-0.06}$ |
| 0.05 | — | $1.9^{+0.9}_{-0.4}$ | $0.329^{+0.010}_{-0.011}$ | $0.18^{+0.05}_{-0.05}$ |
| 0.10 | — | $1.6^{+0.9}_{-0.4}$ | $0.303^{+0.011}_{-0.012}$ | $0.18^{+0.05}_{-0.05}$ |
| 0.15 | — | $1.6^{+0.5}_{-0.3}$ | $0.285^{+0.007}_{-0.007}$ | $0.18^{+0.04}_{-0.03}$ |
| 0.20 | — | $1.6^{+0.5}_{-0.3}$ | $0.262^{+0.007}_{-0.006}$ | $0.18^{+0.03}_{-0.03}$ |
| 0.25 | — | $1.7^{+0.3}_{-0.3}$ | $0.234^{+0.005}_{-0.005}$ | $0.18^{+0.03}_{-0.02}$ |
| 0.30 | — | $1.7^{+0.6}_{-0.4}$ | $0.195^{+0.008}_{-0.008}$ | $0.19^{+0.03}_{-0.04}$ |
| 0.35 | — | $1.7^{+0.2}_{-0.2}$ | $0.154^{+0.005}_{-0.004}$ | $0.19^{+0.02}_{-0.02}$ |
| 0.40 | $0.09^{+0.04}_{-0.04}$ | $1.56^{+0.16}_{-0.14}$ | $0.112^{+0.002}_{-0.002}$ | $0.204^{+0.015}_{-0.015}$ |
| 0.41 | $0.10^{+0.05}_{-0.05}$ | $1.5^{+0.2}_{-0.2}$ | $0.098^{+0.004}_{-0.003}$ | $0.208^{+0.015}_{-0.016}$ |
| 0.42 | $0.12^{+0.10}_{-0.08}$ | $1.5^{+0.4}_{-0.3}$ | $0.089^{+0.004}_{-0.005}$ | $0.21^{+0.03}_{-0.03}$ |
| 0.43 | $0.12^{+0.07}_{-0.06}$ | $1.5^{+0.2}_{-0.3}$ | $0.070^{+0.005}_{-0.005}$ | $0.21^{+0.03}_{-0.02}$ |
| 0.44 | $0.10^{+0.07}_{-0.07}$ | $1.4^{+0.3}_{-0.2}$ | $0.053^{+0.005}_{-0.005}$ | $0.21^{+0.03}_{-0.03}$ |
| 0.45 | $0.07^{+0.17}_{-0.10}$ | $1.4^{+0.6}_{-0.4}$ | $0.025^{+0.010}_{-0.010}$ | $0.21^{+0.06}_{-0.06}$ |
| 0.46 | $0.21^{+0.09}_{-0.07}$ | $1.3^{+0.3}_{-0.3}$ | $0.005^{+0.003}_{-0.003}$ | $0.22^{+0.04}_{-0.04}$ |
| 0.47 | $0.34^{+0.06}_{-0.05}$ | $1.22^{+0.19}_{-0.15}$ | $0.0008^{+0.0007}_{-0.0008}$ | $0.22^{+0.03}_{-0.03}$ |
| 0.48 | $0.45^{+0.06}_{-0.04}$ | $1.3^{+0.2}_{-0.2}$ | $0.00012^{+0.00012}_{-0.00013}$ | $0.21^{+0.04}_{-0.04}$ |
| 0.49 | $0.59^{+0.06}_{-0.03}$ | $1.25^{+0.14}_{-0.12}$ | $0.00001^{+0.00003}_{-0.00002}$ | $0.21^{+0.03}_{-0.02}$ |
| 0.50 | $0.70^{+0.04}_{-0.04}$ | $1.35^{+0.15}_{-0.12}$ | — | $0.21^{+0.02}_{-0.03}$ |
| 0.525 | $1.18^{+0.04}_{-0.03}$ | $0.98^{+0.11}_{-0.10}$ | — | $0.233^{+0.019}_{-0.019}$ |
| 0.55 | $1.79^{+0.01}_{-0.02}$ | $0.40^{+0.06}_{-0.05}$ | — | $0.30^{+0.02}_{-0.01}$ |
| 0.575 | $2.06^{+0.03}_{-0.03}$ | $0.39^{+0.11}_{-0.10}$ | — | $0.288^{+0.019}_{-0.018}$ |
| 0.60 | $2.34^{+0.05}_{-0.05}$ | $0.33^{+0.17}_{-0.15}$ | — | $0.28^{+0.02}_{-0.03}$ |
| 0.65 | $2.97^{+0.04}_{-0.04}$ | — | — | $0.297^{+0.016}_{-0.016}$ |
| 0.70 | $3.41^{+0.05}_{-0.04}$ | — | — | $0.254^{+0.015}_{-0.016}$ |
| 0.75 | $3.73^{+0.08}_{-0.06}$ | — | — | $0.20^{+0.02}_{-0.02}$ |
| 0.80 | $4.02^{+0.05}_{-0.05}$ | — | — | $0.155^{+0.011}_{-0.011}$ |

Table 4: The results from fitting $C_\mathrm{s}$ (left columns) and $C_\mathrm{d}$ (right columns) to (7c) for $16^3$.

| $\beta$ | $N = 6$ | | | $N = 8$ | | |
|---|---|---|---|---|---|---|
| | # | $\Lambda$ | $d_{\mathrm{f}}$ | # | $\Lambda$ | $d_{\mathrm{f}}$ |
| 0.00 | 4.48(6) | 58.2(1.1) | 1.2207(14) | 8.37(8) | 68.4(9) | 1.2208(9) |
| 0.05 | 4.48(5) | 56.3(1.1) | 1.2079(12) | 8.57(11) | 65.2(9) | 1.2057(8) |
| 0.10 | 4.66(6) | 51.2(9) | 1.1929(11) | 8.73(8) | 60.6(7) | 1.1912(8) |
| 0.15 | 4.77(7) | 49.1(1.2) | 1.1767(12) | 9.12(8) | 55.6(7) | 1.1760(10) |
| 0.20 | 4.88(6) | 45.0(9) | 1.1615(12) | 9.30(9) | 52.7(7) | 1.1614(8) |
| 0.25 | 5.06(6) | 41.1(7) | 1.1490(12) | 9.76(9) | 47.3(7) | 1.1471(8) |
| 0.30 | 5.18(7) | 38.1(7) | 1.1314(10) | 10.00(10) | 43.3(6) | 1.1333(6) |
| 0.35 | 5.48(8) | 33.0(8) | 1.1167(12) | 10.84(11) | 36.6(5) | 1.1178(7) |
| 0.40 | 6.07(9) | 25.6(6) | 1.0984(12) | 11.91(13) | 29.4(4) | 1.1011(7) |
| 0.41 | 6.16(11) | 24.7(1.0) | 1.0945(17) | 12.20(17) | 27.9(5) | 1.0976(7) |
| 0.42 | 6.63(12) | 23.3(1.3) | 1.0866(12) | 12.91(17) | 25.3(5) | 1.0940(7) |
| 0.43 | 6.76(10) | 19.6(7) | 1.0811(12) | 13.43(18) | 23.1(5) | 1.0885(8) |
| 0.44 | 6.87(9) | 17.5(6) | 1.0774(13) | 14.51(2) | 19.4(5) | 1.0821(10) |
| 0.45 | 7.11(7) | 15.2(4) | 1.0708(13) | 15.4(2) | 16.2(5) | 1.0750(10) |
| 0.46 | 7.06(8) | 13.5(5) | 1.0639(14) | 16.0(2) | 13.6(4) | 1.0677(12) |
| 0.47 | 6.92(13) | 12.6(5) | 1.0609(11) | 16.18(13) | 11.4(2) | 1.0611(8) |
| 0.48 | 6.92(7) | 10.7(4) | 1.0545(17) | 15.87(12) | 10.4(3) | 1.0564(12) |
| 0.49 | 6.80(7) | 9.3(2) | 1.0486(15) | 15.49(10) | 9.01(14) | 1.0492(9) |
| 0.50 | 6.25(8) | 8.45(15) | 1.0432(14) | 14.65(12) | 8.34(13) | 1.0449(9) |
| 0.525 | 5.35(8) | 7.19(10) | 1.0292(11) | 12.35(14) | 6.97(7) | 1.0343(8) |
| 0.55 | 4.37(8) | 6.41(8) | 1.0229(13) | 10.14(11) | 6.37(6) | 1.0261(8) |
| 0.575 | 3.26(8) | 5.72(7) | 1.0164(10) | 7.91(14) | 5.91(4) | 1.0196(8) |
| 0.60 | 2.60(6) | 5.21(7) | 1.0105(8) | 6.30(8) | 5.64(5) | 1.0151(7) |
| 0.65 | 1.51(4) | 3.86(9) | 1.0035(6) | 3.69(6) | 4.95(5) | 1.0076(6) |
| 0.70 | 0.89(4) | 2.70(9) | 1.0020(6) | 2.08(4) | 4.15(4) | 1.0051(10) |
| 0.75 | 0.52(3) | 1.81(9) | 1.0011(4) | 1.17(4) | 3.16(7) | 1.0017(4) |
| 0.80 | 0.274(18) | 1.05(7) | 1.0005(3) | 0.66(3) | 2.17(8) | 1.0009(3) |
| 0.85 | 0.165(12) | 0.65(4) | 1.0002(3) | 0.37(2) | 1.33(7) | 1.0003(2) |
| 0.90 | 0.119(13) | 0.45(4) | 1.0(−) | 0.232(18) | 0.85(7) | 1.00014(15) |
| 0.95 | 0.053(7) | 0.21(3) | 1.0(−) | 0.120(16) | 0.42(5) | 1.0(−) |
| 1.00 | 0.035(7) | 0.14(3) | 1.0(−) | 0.068(12) | 0.23(4) | 1.0(−) |

Table 5: The results for #, $\Lambda$, and $d_{\mathrm{f}}$ for $N = 6$ and $N = 8$.

|  | $N = 12$ | | | $N = 16$ | | |
|---|---|---|---|---|---|---|
| $\beta$ | # | $\Lambda$ | $d_{\mathrm{f}}$ | # | $\Lambda$ | $d_{\mathrm{f}}$ |
| 0.00 | 23.89(16) | 75.2(6) | 1.2208(4) | 54.2(2) | 76.9(4) | 1.2209(3) |
| 0.05 | 24.61(16) | 70.6(5) | 1.2065(5) | 55.7(2) | 72.4(4) | 1.2065(3) |
| 0.10 | 25.97(18) | 64.8(5) | 1.1919(5) | 58.6(3) | 66.3(3) | 1.1912(3) |
| 0.15 | 26.84(16) | 59.8(3) | 1.1768(4) | 60.4(2) | 61.8(3) | 1.1771(3) |
| 0.20 | 27.75(18) | 55.5(4) | 1.1625(4) | 63.2(2) | 56.4(3) | 1.1625(3) |
| 0.25 | 29.27(16) | 49.7(4) | 1.1479(4) | 66.4(3) | 50.9(2) | 1.1484(3) |
| 0.30 | 30.54(17) | 44.6(3) | 1.1333(3) | 70.0(3) | 45.3(2) | 1.1335(3) |
| 0.35 | 32.8(3) | 37.9(3) | 1.1176(4) | 74.8(3) | 39.02(18) | 1.1184(3) |
| 0.40 | 35.8(2) | 31.2(3) | 1.1022(4) | 82.0(5) | 31.82(16) | 1.1027(3) |
| 0.41 | 36.6(3) | 29.5(3) | 1.0984(4) | 84.3(4) | 29.91(15) | 1.0990(3) |
| 0.42 | 37.8(3) | 27.5(2) | 1.0942(4) | 86.2(3) | 28.38(15) | 1.0962(3) |
| 0.43 | 39.6(4) | 25.3(3) | 1.0916(4) | 89.6(5) | 26.1(2) | 1.0918(3) |
| 0.44 | 41.2(6) | 23.2(5) | 1.0865(4) | 94.0(6) | 23.7(2) | 1.0879(3) |
| 0.45 | 45.2(5) | 19.3(3) | 1.0813(5) | 103.8(1.0) | 19.7(3) | 1.0821(4) |
| 0.46 | 49.5(6) | 15.5(4) | 1.0742(6) | 117.4(9) | 15.1(2) | 1.0740(4) |
| 0.47 | 53.3(3) | 11.7(3) | 1.0638(7) | 124.3(5) | 12.32(12) | 1.0670(4) |
| 0.48 | 53.0(3) | 10.55(14) | 1.0599(6) | 125.3(4) | 10.61(8) | 1.0603(4) |
| 0.49 | 51.7(2) | 9.24(8) | 1.0526(6) | 122.6(3) | 9.48(6) | 1.0544(3) |
| 0.50 | 49.6(2) | 8.52(2) | 1.0488(5) | 117.4(4) | 8.53(4) | 1.0488(3) |
| 0.525 | 42.2(3) | 7.17(4) | 1.0368(5) | 100.3(3) | 7.21(3) | 1.0380(4) |
| 0.55 | 34.0(3) | 6.42(3) | 1.0281(5) | 81.8(3) | 6.480(17) | 1.0295(3) |
| 0.575 | 27.2(2) | 5.90(3) | 1.0212(5) | 64.1(3) | 5.953(17) | 1.0223(3) |
| 0.60 | 21.2(2) | 5.62(3) | 1.0175(5) | 50.3(4) | 5.604(15) | 1.0176(3) |
| 0.65 | 12.97(18) | 5.21(2) | 1.0104(4) | 30.1(2) | 5.195(14) | 1.0107(3) |
| 0.70 | 7.40(13) | 4.92(2) | 1.0050(5) | 17.5(2) | 4.922(13) | 1.0062(3) |
| 0.75 | 4.40(7) | 4.68(3) | 1.0026(3) | 10.30(15) | 4.750(19) | 1.0041(3) |
| 0.80 | 2.63(6) | 4.30(6) | 1.0019(3) | 6.06(8) | 4.670(19) | 1.0024(3) |
| 0.85 | 1.52(4) | 3.40(7) | 1.0007(3) | 3.61(5) | 4.41(3) | 1.0009(2) |
| 0.90 | 0.70(3) | 2.22(8) | 1.0005(2) | 2.06(4) | 3.76(7) | 1.0007(2) |
| 0.95 | 0.54(3) | 1.68(7) | 1.0003(2) | 1.20(3) | 3.03(8) | 1.0003(3) |
| 1.00 | 0.31(2) | 1.03(6) | 1.0004(3) | 0.73(3) | 2.18(8) | 1.00012(11) |

Table 6: The results for #, $\Lambda$, and $d_{\mathrm{f}}$ for $N = 12$ and $N = 16$.

| | N = 6 | | | | N = 8 | | | |
|---|---|---|---|---|---|---|---|---|
| $\beta$ | $a$ | $b$ | $c$ | $d(\cdot 10^3)$ | $a$ | $b$ | $c$ | $d(\cdot 10^4)$ |
| 0.00 | $0.00^{+0.10}_{-0.02}$ | $2.3^{+0.5}_{-0.2}$ | $1.6^{+0.8}_{-0.8}$ | $3.4^{+1.9}_{-1.8}$ | $0.00^{+0.09}_{-0.01}$ | $2.4^{+0.4}_{-0.2}$ | $1.3^{+0.7}_{-0.6}$ | $0.9^{+0.4}_{-0.5}$ |
| 0.05 | $0.00^{+0.15}_{-0.02}$ | $2.4^{+0.5}_{-0.3}$ | $1.6^{+0.8}_{-0.8}$ | $4^{+2}_{-2}$ | $0.00^{+0.02}_{-0.02}$ | $2.4^{+0.6}_{-0.3}$ | $1.3^{+0.6}_{-0.6}$ | $0.9^{+0.6}_{-0.6}$ |
| 0.10 | $0.00^{+0.10}_{-0.02}$ | $2.3^{+0.3}_{-0.3}$ | $1.6^{+0.8}_{-0.8}$ | $3.2^{+1.6}_{-1.6}$ | $0.01^{+0.12}_{-0.12}$ | $2.3^{+0.5}_{-0.2}$ | $1.4^{+0.6}_{-0.7}$ | $1.0^{+0.5}_{-0.5}$ |
| 0.15 | $0.00^{+0.11}_{-0.02}$ | $2.3^{+0.5}_{-0.2}$ | $1.6^{+0.9}_{-0.8}$ | $3.5^{+2.0}_{-2.0}$ | $0.01^{+0.12}_{-0.12}$ | $2.4^{+0.4}_{-0.3}$ | $1.4^{+0.6}_{-0.7}$ | $1.0^{+0.5}_{-0.5}$ |
| 0.20 | $0.00^{+0.02}_{-0.01}$ | $2.3^{+0.4}_{-0.2}$ | $1.6^{+0.9}_{-0.9}$ | $3.4^{+2.0}_{-2.0}$ | $0.00^{+0.02}_{-0.02}$ | $2.4^{+0.4}_{-0.2}$ | $1.3^{+0.6}_{-0.6}$ | $1.1^{+0.5}_{-0.6}$ |
| 0.25 | $0.00^{+0.06}_{-0.07}$ | $2.2^{+0.4}_{-0.2}$ | $1.6^{+0.9}_{-0.9}$ | $3.1^{+1.6}_{-1.7}$ | $0.00^{+0.08}_{-0.02}$ | $2.4^{+0.4}_{-0.1}$ | $1.4^{+0.6}_{-0.6}$ | $1.2^{+0.4}_{-0.4}$ |
| 0.30 | $0.00^{+0.06}_{-0.02}$ | $2.2^{+0.4}_{-0.1}$ | $1.6^{+0.9}_{-0.9}$ | $3.7^{+1.5}_{-1.5}$ | $0.00^{+0.07}_{-0.01}$ | $2.3^{+0.4}_{-0.2}$ | $1.4^{+0.7}_{-0.6}$ | $1.0^{+0.5}_{-0.5}$ |
| 0.35 | $0.00^{+0.06}_{-0.02}$ | $2.2^{+0.3}_{-0.2}$ | $1.5^{+0.9}_{-1.0}$ | $3.4^{+1.8}_{-1.8}$ | $0.00^{+0.07}_{-0.01}$ | $2.3^{+0.3}_{-0.2}$ | $1.4^{+0.6}_{-0.7}$ | $1.1^{+0.3}_{-0.4}$ |
| 0.40 | $0.00^{+0.03}_{-0.02}$ | $2.1^{+0.2}_{-0.2}$ | $1.2^{+0.9}_{-0.9}$ | $3.7^{+1.3}_{-1.3}$ | $0.00^{+0.05}_{-0.02}$ | $2.2^{+0.3}_{-0.1}$ | $1.4^{+0.7}_{-0.7}$ | $1.1^{+0.3}_{-0.3}$ |
| 0.41 | $0.00^{+0.03}_{-0.02}$ | $2.06^{+0.17}_{-0.13}$ | $1.1^{+0.9}_{-0.9}$ | $3.6^{+1.3}_{-1.3}$ | $0.00^{+0.04}_{-0.02}$ | $2.3^{+0.2}_{-0.2}$ | $1.3^{+0.7}_{-0.7}$ | $1.1^{+0.5}_{-0.5}$ |
| 0.42 | $0.00^{+0.01}_{-0.02}$ | $2.03^{+0.17}_{-0.11}$ | $0.7^{+0.8}_{-0.8}$ | $3.7^{+1.3}_{-1.3}$ | $0.00^{+0.05}_{-0.02}$ | $2.2^{+0.2}_{-0.2}$ | $1.2^{+0.8}_{-0.7}$ | $1.1^{+0.3}_{-0.5}$ |
| 0.43 | $0.00^{+0.01}_{-0.02}$ | $2.02^{+0.14}_{-0.10}$ | $0.7^{+0.8}_{-0.8}$ | $3.8^{+1.1}_{-1.2}$ | $0.00^{+0.03}_{-0.02}$ | $2.17^{+0.17}_{-0.12}$ | $1.2^{+0.8}_{-0.5}$ | $1.1^{+0.3}_{-0.3}$ |
| 0.44 | $0.00^{+0.02}_{-0.03}$ | $2.04^{+0.11}_{-0.08}$ | $0.2^{+0.7}_{-0.7}$ | $4.0^{+1.1}_{-1.1}$ | $0.00^{+0.03}_{-0.02}$ | $2.03^{+0.18}_{-0.12}$ | $1.1^{+0.9}_{-0.7}$ | $0.9^{+0.3}_{-0.3}$ |
| 0.45 | $0.00^{+0.02}_{-0.02}$ | $2.02^{+0.15}_{-0.10}$ | $1.0^{+0.8}_{-0.8}$ | $4.5^{+1.2}_{-1.2}$ | $0.01^{+0.03}_{-0.01}$ | $1.97^{+0.19}_{-0.13}$ | $1.2^{+0.7}_{-0.8}$ | $0.9^{+0.4}_{-0.3}$ |
| 0.46 | $0.01^{+0.03}_{-0.02}$ | $1.98^{+0.19}_{-0.12}$ | $0.9^{+0.8}_{-0.8}$ | $4.5^{+1.5}_{-1.6}$ | $0.01^{+0.03}_{-0.01}$ | $1.99^{+0.13}_{-0.13}$ | $1.2^{+0.8}_{-0.8}$ | $1.1^{+0.3}_{-0.3}$ |
| 0.47 | $0.01^{+0.03}_{-0.01}$ | $1.95^{+0.17}_{-0.12}$ | $1.0^{+0.9}_{-0.8}$ | $4.3^{+1.3}_{-1.3}$ | $0.02^{+0.02}_{-0.01}$ | $1.88^{+0.11}_{-0.08}$ | $1.0^{+0.9}_{-0.8}$ | $0.97^{+0.19}_{-0.19}$ |
| 0.48 | $0.04^{+0.03}_{-0.02}$ | $1.77^{+0.15}_{-0.11}$ | $1.2^{+0.8}_{-0.9}$ | $4.0^{+1.1}_{-1.1}$ | $0.023^{+0.020}_{-0.012}$ | $1.88^{+0.11}_{-0.08}$ | $1.0^{+0.9}_{-0.9}$ | $0.99^{+0.18}_{-0.19}$ |
| 0.49 | $0.04^{+0.03}_{-0.02}$ | $1.82^{+0.17}_{-0.11}$ | $1.0^{+0.9}_{-0.9}$ | $4.3^{+1.1}_{-1.0}$ | $0.041^{+0.019}_{-0.013}$ | $1.81^{+0.09}_{-0.08}$ | $1.0^{+0.8}_{-0.8}$ | $0.99^{+0.15}_{-0.15}$ |
| 0.50 | $0.07^{+0.03}_{-0.02}$ | $1.62^{+0.11}_{-0.09}$ | $1.1^{+0.9}_{-0.8}$ | $3.5^{+0.6}_{-0.6}$ | $0.05^{+0.02}_{-0.01}$ | $1.76^{+0.10}_{-0.10}$ | $1.0^{+1.0}_{-1.0}$ | $0.94^{+0.16}_{-0.17}$ |
| 0.525 | $0.15^{+0.03}_{-0.03}$ | $1.16^{+0.10}_{-0.09}$ | $1.1^{+0.8}_{-0.8}$ | $2.3^{+0.4}_{-0.3}$ | $0.111^{+0.019}_{-0.016}$ | $1.53^{+0.09}_{-0.07}$ | $1.0^{+0.8}_{-0.8}$ | $0.79^{+0.11}_{-0.12}$ |
| 0.55 | $0.24^{+0.03}_{-0.02}$ | $0.86^{+0.18}_{-0.11}$ | $1.2^{+0.8}_{-0.9}$ | $1.9^{+0.3}_{-0.4}$ | $0.15^{+0.02}_{-0.02}$ | $1.48^{+0.08}_{-0.06}$ | $0.9^{+0.9}_{-1.0}$ | $0.74^{+0.11}_{-0.10}$ |
| 0.575 | $0.32^{+0.04}_{-0.02}$ | $0.59^{+0.12}_{-0.10}$ | $1.2^{+0.8}_{-0.8}$ | $1.5^{+0.2}_{-0.3}$ | $0.258^{+0.018}_{-0.016}$ | $0.95^{+0.06}_{-0.06}$ | $1.0^{+0.9}_{-0.9}$ | $0.46^{+0.04}_{-0.05}$ |
| 0.60 | $0.45^{+0.03}_{-0.03}$ | $0.10^{+0.11}_{-0.09}$ | $1.2^{+0.7}_{-0.7}$ | $1.05^{+0.17}_{-0.17}$ | $0.31^{+0.02}_{-0.02}$ | $0.80^{+0.07}_{-0.07}$ | $1.0^{+0.8}_{-0.8}$ | $0.39^{+0.04}_{-0.05}$ |
| 0.65 | $0.57^{+0.03}_{-0.03}$ | $0.00^{+0.11}_{-0.10}$ | $0.9^{+0.7}_{-0.7}$ | $0.95^{+0.15}_{-0.14}$ | $0.58^{+0.03}_{-0.03}$ | $0.01^{+0.10}_{-0.07}$ | $1.0^{+1.0}_{-1.0}$ | $0.25^{+0.04}_{-0.03}$ |

Table 7: The results for fitting $\rho(l)$ to (7c).

| | $N = 12$ | | | | $N = 16$ | | | |
|---|---|---|---|---|---|---|---|---|
| $\beta$ | $a$ | $b$ | $c$ | $d(\cdot 10^4)$ | $a$ | $b$ | $c$ | $d(\cdot 10^4)$ |
| 0.00 | $0.0^{+0.1}_{-0.1}$ | $2.4^{+0.4}_{-0.2}$ | $1.2^{+0.8}_{-0.8}$ | $3.3^{+1.6}_{-1.7}$ | $0.01^{+0.08}_{-0.01}$ | $2.4^{+0.3}_{-0.2}$ | $1.1^{+0.9}_{-0.9}$ | $7^{+3}_{-3}$ |
| 0.05 | $0.06^{+0.06}_{-0.07}$ | $2.4^{+0.5}_{-0.2}$ | $1.2^{+0.8}_{-0.8}$ | $3.4^{+1.9}_{-1.9}$ | $0.01^{+0.06}_{-0.02}$ | $2.41^{+0.20}_{-0.15}$ | $1.1^{+0.9}_{-0.9}$ | $8^{+3}_{-3}$ |
| 0.10 | $0.00^{+0.10}_{-0.06}$ | $2.5^{+0.3}_{-0.2}$ | $1.2^{+0.8}_{-0.8}$ | $3.7^{+1.2}_{-1.5}$ | $0.01^{+0.06}_{-0.01}$ | $2.4^{+0.3}_{-0.2}$ | $1.1^{+0.9}_{-0.9}$ | $8^{+3}_{-2}$ |
| 0.15 | $0.01^{+0.09}_{-0.02}$ | $2.4^{+0.3}_{-0.2}$ | $1.2^{+0.8}_{-0.7}$ | $3.4^{+1.4}_{-1.4}$ | $0.01^{+0.07}_{-0.01}$ | $2.3^{+0.3}_{-0.2}$ | $1.2^{+0.8}_{-0.8}$ | $8^{+3}_{-2}$ |
| 0.20 | $0.01^{+0.06}_{-0.02}$ | $2.4^{+0.2}_{-0.2}$ | $1.2^{+0.8}_{-0.7}$ | $3.5^{+1.4}_{-1.4}$ | $0.01^{+0.06}_{-0.01}$ | $2.3^{+0.3}_{-0.2}$ | $1.1^{+0.9}_{-0.8}$ | $8^{+3}_{-3}$ |
| 0.25 | $0.02^{+0.06}_{-0.01}$ | $2.3^{+0.2}_{-0.2}$ | $1.2^{+0.8}_{-0.7}$ | $3.5^{+1.2}_{-1.1}$ | $0.01^{+0.05}_{-0.01}$ | $2.30^{+0.18}_{-0.12}$ | $1.1^{+0.9}_{-0.8}$ | $8^{+2}_{-2}$ |
| 0.30 | $0.01^{+0.06}_{-0.01}$ | $2.3^{+0.2}_{-0.2}$ | $1.2^{+0.8}_{-0.7}$ | $3.5^{+1.2}_{-1.3}$ | $0.01^{+0.05}_{-0.01}$ | $2.26^{+0.20}_{-0.14}$ | $1.2^{+0.8}_{-0.9}$ | $8^{+3}_{-3}$ |
| 0.35 | $0.02^{+0.07}_{-0.02}$ | $2.2^{+0.3}_{-0.2}$ | $1.3^{+0.7}_{-0.7}$ | $3.3^{+1.6}_{-1.5}$ | $0.02^{+0.04}_{-0.02}$ | $2.21^{+0.20}_{-0.11}$ | $1.1^{+0.9}_{-0.8}$ | $8^{+3}_{-3}$ |
| 0.40 | $0.02^{+0.04}_{-0.02}$ | $2.17^{+0.19}_{-0.13}$ | $1.3^{+0.6}_{-1.0}$ | $3.5^{+1.0}_{-1.0}$ | $0.02^{+0.04}_{-0.02}$ | $2.18^{+0.20}_{-0.15}$ | $1.2^{+0.8}_{-0.9}$ | $8^{+3}_{-3}$ |
| 0.41 | $0.02^{+0.03}_{-0.02}$ | $2.14^{+0.17}_{-0.12}$ | $1.3^{+0.6}_{-0.7}$ | $3.4^{+0.9}_{-0.9}$ | $0.01^{+0.04}_{-0.02}$ | $2.15^{+0.19}_{-0.12}$ | $1.2^{+0.8}_{-0.8}$ | $8^{+2}_{-3}$ |
| 0.42 | $0.01^{+0.05}_{-0.01}$ | $2.1^{+0.3}_{-0.1}$ | $1.3^{+0.7}_{-0.7}$ | $3.5^{+1.0}_{-1.0}$ | $0.01^{+0.03}_{-0.01}$ | $2.15^{+0.14}_{-0.11}$ | $1.2^{+0.8}_{-0.8}$ | $8.2^{+1.7}_{-1.8}$ |
| 0.43 | $0.01^{+0.04}_{-0.01}$ | $2.1^{+0.2}_{-0.2}$ | $1.3^{+0.7}_{-0.7}$ | $3.5^{+1.2}_{-1.3}$ | $0.01^{+0.03}_{-0.01}$ | $2.14^{+0.14}_{-0.10}$ | $1.2^{+0.8}_{-0.8}$ | $8.2^{+1.9}_{-1.9}$ |
| 0.44 | $0.01^{+0.05}_{-0.01}$ | $2.1^{+0.3}_{-0.2}$ | $1.3^{+0.6}_{-0.6}$ | $3.4^{+1.5}_{-1.5}$ | $0.01^{+0.03}_{-0.01}$ | $2.13^{+0.16}_{-0.12}$ | $1.2^{+0.8}_{-0.8}$ | $8^{+2}_{-2}$ |
| 0.45 | $0.01^{+0.03}_{-0.01}$ | $2.03^{+0.18}_{-0.12}$ | $1.3^{+0.7}_{-0.8}$ | $3.3^{+1.0}_{-1.0}$ | $0.01^{+0.03}_{-0.01}$ | $2.08^{+0.20}_{-0.13}$ | $1.2^{+0.8}_{-0.8}$ | $8^{+3}_{-3}$ |
| 0.46 | $0.01^{+0.02}_{-0.01}$ | $2.01^{+0.12}_{-0.09}$ | $1.1^{+0.8}_{-0.8}$ | $3.4^{+0.9}_{-0.9}$ | $0.01^{+0.02}_{-0.01}$ | $2.01^{+0.11}_{-0.08}$ | $1.1^{+0.8}_{-0.8}$ | $8.0^{+1.8}_{-1.8}$ |
| 0.47 | $0.015^{+0.013}_{-0.008}$ | $1.93^{+0.09}_{-0.08}$ | $1.1^{+0.8}_{-0.8}$ | $3.4^{+0.6}_{-0.6}$ | $0.01^{+0.01}_{-0.01}$ | $1.96^{+0.06}_{-0.05}$ | $1.1^{+0.9}_{-0.9}$ | $8.0^{+1.0}_{-1.0}$ |
| 0.48 | $0.021^{+0.012}_{-0.007}$ | $1.87^{+0.06}_{-0.06}$ | $1.1^{+0.8}_{-0.8}$ | $3.2^{+0.3}_{-0.3}$ | $0.015^{+0.008}_{-0.005}$ | $1.93^{+0.04}_{-0.05}$ | $1.0^{+0.9}_{-0.9}$ | $8.0^{+0.8}_{-0.8}$ |
| 0.49 | $0.031^{+0.011}_{-0.008}$ | $1.86^{+0.05}_{-0.05}$ | $1.1^{+0.8}_{-0.8}$ | $3.4^{+0.3}_{-0.3}$ | $0.024^{+0.007}_{-0.005}$ | $1.88^{+0.03}_{-0.04}$ | $1.1^{+0.9}_{-1.0}$ | $7.8^{+0.5}_{-0.5}$ |
| 0.50 | $0.040^{+0.010}_{-0.009}$ | $1.82^{+0.05}_{-0.04}$ | $1.0^{+0.8}_{-0.9}$ | $3.2^{+0.3}_{-0.2}$ | $0.036^{+0.007}_{-0.005}$ | $1.84^{+0.03}_{-0.03}$ | $1.0^{+0.9}_{-0.9}$ | $7.7^{+0.4}_{-0.4}$ |
| 0.525 | $0.078^{+0.012}_{-0.010}$ | $1.72^{+0.05}_{-0.04}$ | $1.1^{+1.0}_{-0.5}$ | $3.1^{+0.3}_{-0.3}$ | $0.073^{+0.007}_{-0.007}$ | $1.72^{+0.03}_{-0.03}$ | $1.0^{+1.0}_{-1.0}$ | $7.0^{+0.4}_{-0.4}$ |
| 0.55 | $0.131^{+0.012}_{-0.010}$ | $1.55^{+0.05}_{-0.04}$ | $1.1^{+0.9}_{-0.9}$ | $2.60^{+0.19}_{-0.19}$ | $0.121^{+0.006}_{-0.006}$ | $1.56^{+0.02}_{-0.02}$ | $1.1^{+1.0}_{-1.0}$ | $5.96^{+0.20}_{-0.20}$ |
| 0.575 | $0.221^{+0.013}_{-0.011}$ | $1.15^{+0.04}_{-0.04}$ | $1.2^{+1.0}_{-1.0}$ | $1.81^{+0.12}_{-0.13}$ | $0.206^{+0.006}_{-0.006}$ | $1.19^{+0.02}_{-0.02}$ | $0.8^{+0.8}_{-1.0}$ | $4.28^{+0.16}_{-0.15}$ |
| 0.60 | $0.258^{+0.013}_{-0.014}$ | $1.14^{+0.05}_{-0.04}$ | $1.2^{+1.0}_{-1.0}$ | $1.69^{+0.12}_{-0.12}$ | $0.243^{+0.013}_{-0.010}$ | $1.24^{+0.04}_{-0.04}$ | $0.9^{+1.0}_{-0.9}$ | $4.4^{+0.3}_{-0.3}$ |
| 0.65 | $0.430^{+0.013}_{-0.013}$ | $0.46^{+0.05}_{-0.04}$ | $0.9^{+0.8}_{-0.8}$ | $0.86^{+0.06}_{-0.06}$ | $0.464^{+0.007}_{-0.010}$ | $0.240^{+0.024}_{-0.024}$ | $1.5^{+1.3}_{-1.3}$ | $1.71^{+0.07}_{-0.07}$ |

Table 8: The results for fitting $\rho(l)$ to (7c).

| $\beta$ | N=6 | N=8 | N=12 | N=16 |
|---|---|---|---|---|
| 0.00 | 234(7) | 527(11) | 1660(30) | 3870(60) |
| 0.05 | 229(9) | 504(10) | 1640(30) | 3750(60) |
| 0.10 | 221(6) | 488(16) | 1561(19) | 3540(60) |
| 0.15 | 212(8) | 471(15) | 1490(30) | 3410(50) |
| 0.20 | 204(6) | 452(12) | 1410(30) | 3240(40) |
| 0.25 | 199(7) | 421(12) | 1330(30) | 3020(50) |
| 0.30 | 182(5) | 399(15) | 1240(30) | 2800(60) |
| 0.35 | 173(6) | 365(11) | 1130(30) | 2530(50) |
| 0.40 | 151(9) | 328(20) | 960(30) | 2160(70) |
| 0.41 | 158(14) | 317(18) | 940(40) | 2030(80) |
| 0.42 | 150(8) | 306(14) | 880(40) | 1950(50) |
| 0.43 | 139(9) | 300(20) | 850(40) | 1830(60) |
| 0.44 | 137(11) | 272(12) | 811(13) | 1700(50) |
| 0.45 | 128(13) | 250(30) | 710(50) | 1460(150) |
| 0.46 | 123(14) | 210(30) | 580(90) | 1040(200) |
| 0.47 | 118(17) | 170(40) | 350(90) | 610(170) |
| 0.48 | 95(16) | 140(30) | 340(80) | 390(50) |
| 0.49 | 80(16) | 120(30) | 230(60) | 230(30) |
| 0.50 | 65(11) | 90(20) | 170(40) | 180(40) |
| 0.525 | 45(9) | 62(19) | 90(20) | 91(11) |
| 0.55 | 36(5) | 40(11) | 51(6) | 61(9) |
| 0.575 | 26(5) | 30(10) | 46(10) | 48(6) |
| 0.60 | 21(3) | 27(6) | 33(12) | 43(11) |
| 0.65 | 14(3) | 20(5) | 23(4) | 22.4(1.5) |
| 0.70 | 11(3) | 13.2(1.6) | 16(3) | 17.6(1.2) |
| 0.75 | 10(2) | 11.8(2.0) | 13(2) | 17(3) |
| 0.80 | 7.6(1.2) | 8.8(1.3) | 10.4(1.2) | 13(2) |
| 0.85 | 7.2(1.8) | 7.2(1.6) | 9.4(1.3) | 12(3) |
| 0.90 | 6.2(1.4) | 6.8(1.0) | 8.6(0.9) | 9.6(1.7) |
| 0.95 | 5.2(1.6) | 6.4(1.5) | 7.4(0.9) | 9.0(1.8) |
| 1.00 | 5.0(1.6) | 5.2(1.3) | 7.0(1.0) | 8.4(1.2) |

Table 9: The results for the average of the largest loop, $\overline{\ell}$.